\documentclass[11pt]{article}

\usepackage{fullpage,microtype}
\usepackage{graphicx,amsfonts,amsmath,amssymb,epsfig,hyperref,color}
\usepackage{algorithm}
\usepackage{array}

\author{
Reut Levi\thanks{Efi Arazi School of Computer Science,  The Interdisciplinary Center, Israel.
  Email: {\tt reut.levi1@idc.ac.il}. This project has received funding from the European Research Council (ERC) under the European Union’s Horizon 2020 research and innovation programme (grant agreement No. 819702). It was also supported by ERC-CoG grant 772839.}
\and
Dana Ron\thanks{School of Electrical Engineering, Tel Aviv University.
  Tel Aviv 69978, Israel.
  Email: {\tt danar@eng.tau.ac.il}. Supported by the Israel Science Foundation grant number nos. 246/08 and  671/13.}
\and
Ronitt Rubinfeld\thanks{CSAIL, MIT.
  Cambridge MA 02139, USA.
Blavatnik School of Computer Science, Tel Aviv University.
  Tel Aviv 69978, Israel.
  Email: {\tt ronitt@csail.mit.edu}. Supported by the Israel Science Foundation grant number no. 1147/09,
nd by NSF grants CCF-1217423 and CCF-1065125.}
}

\newcommand{\eps}{\epsilon}
\newcommand{\eqdef}{\stackrel{\rm def}{=}}
\newcommand{\bitset}{\{0,1\}}

\newtheorem{defn}{Definition}         
\newcommand{\BD}{\begin{defn}} \newcommand{\ED}{\end{defn}}
\newcommand{\BE}{\begin{enumerate}} \newcommand{\EE}{\end{enumerate}}
\newcommand{\BI}{\begin{itemize}} \newcommand{\EI}{\end{itemize}}
\newcommand{\calA}{{\mathcal A}}
\newcommand{\calC}{{\mathcal C}}
\newcommand{\calF}{{\mathcal F}}

\newcommand{\calP}{{\mathcal P}}
\newtheorem{thm}{Theorem}
\newcommand{\BT}{\begin{thm}} \newcommand{\ET}{\end{thm}}
\def\FullBox{\hbox{\vrule width 8pt height 8pt depth 0pt}}
\newcommand{\ourqed}{\;\;\;\FullBox}
\newenvironment{ourproof}{\noindent{\bf Proof:~~}}{\(\ourqed\)}
\newcommand{\BPF}{\begin{ourproof}} \newcommand {\EPF}{\end{ourproof}}
\newenvironment{proofof}[1]{\noindent{\bf Proof of {#1}:~~}}{\(\ourqed\)}
\newcommand{\BPFOF}{\smallskip \begin{proofof}} \newcommand {\EPFOF}{\end{proofof}}
\newcommand{\BEQ}{\begin{equation}} \newcommand{\EEQ}{\end{equation}}
\newcommand{\BEQN}{\begin{eqnarray}}\newcommand{\EEQN}{\end{eqnarray}}
\newtheorem{lem}{Lemma}      
\newcommand{\BL}{\begin{lem}} \newcommand{\EL}{\end{lem}}
\newtheorem{cor}[thm]{Corollary}      
\newcommand{\BC}{\begin{cor}} \newcommand{\EC}{\end{cor}}
\newtheorem{clm}[lem]{Claim}
\newcommand{\BCM}{\begin{clm}} \newcommand{\ECM}{\end{clm}}
\newcommand{\poly}{{\rm poly}}
\newtheorem{fact}{Fact}      
\newcommand{\BF}{\begin{fact}} \newcommand{\EF}{\end{fact}}
\newcommand{\E}{{\rm E}}

\newcommand{\cent}{\sigma}
\newcommand{\cc}{{\rm cc}}
\newcommand{\W}{W}  
\newcommand{\cell}{C}  
\newcommand{\Path}{P}  
\newcommand{\chiW}{\chi_{\mbox{\tiny\it W}}}

\newcommand{\tr}{t}
\renewcommand{\Pr}{\mathrm{Pr}}
\newcommand{\Exp}{\mathrm{Exp}}
\newcommand{\Var}{\mathrm{Var}}
\newcommand{\cov}{\mathrm{Cov}}
\newcommand{\s}[1]{\left\lvert #1 \right\rvert}

\renewcommand{\d}{\Delta} 

\newcommand{\calT}{{\cal T}}
\newcommand{\sz}{k}  
\newcommand{\heavy}{{\cal H}}
\newcommand{\light}{{\cal L}}
\newcommand{\wtd}{{\widetilde{d}}}
\newcommand{\wtG}{{\widetilde{G}}}

\newcommand{\tE}{{\widetilde{E}}}

\newcommand{\z}{\cent}
\newcommand{\x}{x}
\newcommand{\y}{\tau}

\title{Local Algorithms for Sparse Spanning Graphs\footnote{This work is based on part of
the extended abstract~\cite{LRR14} and on the extended abstract~\cite{LRR16}.
Other parts of~\cite{LRR14} have appeared in the journal publication~\cite{LMRRS}.}}

\begin{document}

\begin{titlepage}

\maketitle

\begin{abstract}
Constructing a spanning tree of a graph is
one of the most basic tasks in graph theory.
 We consider a relaxed version of this problem in the setting of local algorithms.
 The relaxation is that the constructed subgraph
 is  a {\em sparse spanning subgraph\/} containing at most $(1+\eps)n$ edges
 (where $n$ is the number of vertices and $\eps$ is a given approximation/sparsity parameter).
 In the local setting, the goal is
to quickly determine whether a given edge $e$ belongs to such a subgraph,
without constructing the whole subgraph, but rather by inspecting (querying)
the local neighborhood of $e$.
The challenge is to maintain consistency. That is, to provide answers
concerning different edges according to
the {\em same\/} spanning subgraph.

 We first show that for general bounded-degree graphs, the query complexity
 of any such algorithm must be $\Omega(\sqrt{n})$. This lower bound holds
 for constant-degree graphs that have high expansion.
 Next we design an algorithm
for (bounded-degree) graphs with high expansion, obtaining a result that roughly
 matches the lower bound.
We then turn to study graphs that exclude a fixed minor (and are hence non-expanding).
 We design an algorithm for such graphs, which may have an unbounded maximum degree.
The query complexity of this
 algorithm is $\poly(1/\eps, h)$ (independent of $n$ and the maximum degree),
 where $h$ is the number of vertices in the excluded minor.

Though our two algorithms are designed for very different types of graphs (and have very different complexities),
on a high-level there are several similarities, and we highlight both the similarities and
the differences.
\end{abstract}

\end{titlepage}

\section{Introduction}
When dealing with large graphs, it is often important to
work with a sparse subgraph that maintains essential
properties, such as  connectivity,
bounded diameter and other distance 
properties (e.g., 
small {\em stretch\/}~\cite{PU89,PS89}), of the original input graph.
Can one provide fast
random access to such a sparsified approximation of the
original input graph?
In this work, we consider the property of connectivity:
Given a connected graph $G=(V,E)$ over $n$ vertices, find a sparse subgraph of $G'$ that spans
$G$.   This task can be accomplished by  constructing a spanning tree in
linear time.
 However, it may be crucial to {\em quickly} determine whether a particular edge $e$ belongs to such a subgraph $G'$, where by ``quickly'' we mean in time much faster than
constructing all of $G'$.   The hope is that by inspecting only some
small local neighborhood of $e$,
one can answer in such a way that maintains consistency with the same $G'$
for all edges $e$. 
We focus on such algorithms (referred to as (centralized) {\em local\/} algorithms~\cite{RTVX}),
which are of use when we do not need to know the answer for every edge at any single point in time,
or if there are several  independent processes that want to determine the answer for edges of their choice, possibly in parallel.

If we insist that $G'$ should have the {\em minimum\/} number of edges
sufficient for spanning $G$, namely, that $G'$ be a spanning tree, then it is easy to
see that the task cannot be performed in general without inspecting almost all of $G$.
To verify this, observe that
 if $G$ consists of a single path, then the algorithm must answer positively on
all edges, while if $G$ consists of a cycle, then the algorithm must answer negatively on
one edge. However, the two cases cannot be distinguished without inspecting a linear
number of edges. On the other hand, suppose we allow the algorithm some more slackness.
Namely,
rather than requiring that $G'$ be a tree, require that it be relatively sparse, i.e., contains at most $(1+\eps)n$ edges,
(allowing it to contain more than $n-1$ edges). Now the algorithm may answer positively on all
cycle edges, so distinguishing between these two cases is no longer necessary.

Given the above observation, the question we consider
is whether a relaxed version of this task can be solved by
inspecting a sublinear number of edges. Specifically, the relaxation is
that the spanning subgraph $G'$ may contain $(1+\eps)\cdot n$ edges,
for a given approximation/sparsity parameter $\eps$.
To be more precise we consider the problem as formulated next.

\BD 
\label{dfn:SSG-alg}
An algorithm $\calA$ is a {\sf Local Sparse Spanning Graph (LSSG) algorithm} if, given
$n\geq 1$, $\eps > 0$,
a sequence of random bits $r\in \{0,1\}^*$
and query access to the incidence-lists representation of a connected graph $G=(V,E)$ over $n$ vertices,\footnote{Namely, the algorithm can query, for any vertex $v\in V$ and index $i$, who is the $i$th neighbor of $v$ (where if $v$ has less than $i$ neighbors, then a special symbol is returned).}
it provides
oracle
 access to a subgraph $G'=(V, E')$ of $G$ such that:
\BE
\item\label{it:connect} $G'$ is connected. 
\item\label{it:internal-rand} $\s{E'} \leq (1+\eps)\cdot n$ with probability at least
    $1-1/\Omega(n)$,
where $E'$ is determined by $G$ and $r$.
\EE
By ``providing oracle access to $G'$'' we mean that
on input $(u,v)\in E$,
    $\calA$ returns whether $(u,v) \in E'$, and for any
    sequence of edges, $\calA$ answers consistently with
    the same $G'$. \\
We consider the following performance measures of the algorithm.
\BI
\item Query complexity: number of graph-queries it makes in order to answer a single oracle-query.
\item Running time: the time it takes to answer a single oracle-query 
(in the word-RAM model).\footnote{In this model (which is attributed to Fredman and Willard~\cite{FW90}), memory is divided into words of length $w$ bits each. Every basic operation on words (in particular, comparison and addition), is assumed to take unit time.  In our setting a word is $\Theta(\log n)$ bits.}
\item Total number of random bits the algorithm uses (i.e. $|r|$). 
\EI
An algorithm $\calA$ is an {\sf LSSG
algorithm for a family of graphs $\mathcal{C}$}
if the above conditions hold, provided that the input graph
$G$ belongs to $\mathcal{C}$.
\ED

We note that the randomness of the algorithm does not affect its query complexity and running time, but only the number of edges in the sparse spanning graph.

The complexity of the algorithm may depend on a bound, $d$, on the maximum degree.
We  note that if $|E| = O(n)$ (i.e., the maximum or average degree is a constant), then clearly one can
get a constant factor approximation (to the minimum number of edges necessary for connectivity) by
simply taking $E'=E$. However, we are interested in a much better approximation of  $(1+\eps)$ for any given
approximation parameter $\eps$.

Two comments are in place regarding the success probability of an LSSG algorithm.
First,  Definition~\ref{dfn:SSG-alg} can be modified so as to allow a higher probability of
 failure in terms of the sparsity (that is, higher than $1/\Omega(n)$ in Item~\ref{it:internal-rand}).
 Second, the definition can be modified so as to allow a probability of failure in terms
 of connectivity (that is, in Item~\ref{it:connect}).
  Since our algorithms ensure the required sparsity with probability $1-1/\Omega(n)$ and always ensure connectivity, for the sake of succinctness, we have chosen not to introduce  additional  parameters in the definition.

We are interested in LSSG algorithms that
for each given edge, perform
as few queries as possible to $G$.
We are also interested in bounding the total amount randomness
used by such algorithms, and their running time (in the word-RAM model).
Note that Item~\ref{it:internal-rand}, which states that the set of edges $E'$ is determined by the graph $G$ and the randomness $r$, implies that the answers of an LSSG algorithm
to queries cannot depend on previously asked queries.

\subsection{Our results}
We start by showing that even after the aforementioned relaxation, local algorithms for sparse spanning
graphs require the inspection of $\Omega(n^{1/2})$ 
edges for each given edge $e$ and for a constant $\eps$.
This lower bound holds for constant-degree graphs with strong expansion properties,
and in our first positive result 
we provide an almost matching upper bound for such graphs.
Namely, the first algorithm we present, the {\sf Sparse-Centers} algorithm\footnote{In the extended abstract~\cite{LRR14}, the algorithm is referred to simply as the {\sf Centers} algorithm.
Here we name it the {\sf Sparse-Centers} algorithm so as to distinguish it from the other (centers based) algorithm that is presented in this work. } 
gives meaningful results for graphs in which the neighborhoods of almost all the vertices in the graph expand in a similar rate.
In particular, for graphs with high expansion we get 
complexity
that grows approximately like $n^{1/2}$. 
More precisely, if the expansion of small sets (of size roughly $O(n^{1/2})$) is $\Omega(d)$, where $d$ is the maximum degree in the graph, then the complexity of the algorithm is $n^{1/2 + O(1/\log d)}$.
The number of random bits the algorithm uses, in terms of $n$, is $\tilde{O}(\sqrt{n})$.~\footnote{We use $\tilde{O}$ for $\tilde{O}(x) = O(x)\cdot \poly(\log x)$.}

We then turn to consider non-expanding graphs.
A family of non-expanding graphs that is of wide interest, is the family of {\em minor-free\/}
graphs, which can be parameterized by the size, $h$, of the excluded minor.
In particular, this family includes planar graphs.
Note that although the average degree in graphs with an excluded fixed minor is constant, the maximum degree might be large as $\Omega(n)$.
Our second algorithm, the {\sf Dense-Centers} algorithm,
is designed for minor-free graphs, and
has complexity polynomial in $h$ and $1/\eps$ (with no dependence on $n$ or on the maximum degree,\footnote{The extension to unbounded degrees does not appear in the extended abstract~\cite{LRR16} and appears only in this version.} so that it works for unbounded-degree graphs).
Another parameter of interest is the {\em stretch}\footnote{Given a connected graph $G=(V, E)$, a subgraph $G'=(V, E')$ is a $t$-spanner of $G$ if for every $u,v \in V$, $\frac{\d_{G'}(u,v)}{\d_{G}(u,v)} \leq t$ (where for a graph $H$ and two vertices $u,v$ in $H$, $\d_H(u,v)$ denotes the distance between $u$ and $v$ in $H$). We refer to $t$  as the {\em stretch factor} of $G'$.} of $G'$~\cite{PU89,PS89}.
The spanning subgraph $G'$ 
has {\em stretch\/}
$\tilde{O}(h/\eps)$.
The length of the random seed the algorithm uses is $\poly(h/\eps)\cdot \log n$.

Observe that while the query complexity of our algorithm for expanding graphs has relatively high dependence on $n$, our algorithm
for minor-free graphs has no dependence on $n$. This raises the natural question regarding the relation between
expansion properties of graphs and the query complexity of the LSSG problem (in terms of the dependence on $n$). This question was addressed in~\cite{LMRRS}, as we further elaborate in Section~\ref{subsec:related}.

\subsection{Our algorithms}\label{subsec:our-algs}
Interestingly, though our two algorithms are designed for very different types of graphs (and have very different complexities),
on a high-level there are several similarities. In particular, underlying each
of the two algorithms is the existence of a (global) partition of the graph vertices
where the partition determines the sparse spanning subgraph, and the local algorithm decides whether  a given edge belongs to the spanning subgraph by constructing parts of the
partition. The differences between the algorithms are in
the way the partition is defined, the precise way the spanning subgraph is determined by the partition (in particular,
the choice of edges between parts)
and the local (partial) construction of the partition.
We further discuss the similarities and differences between the algorithms in Section~\ref{subsubsec:alg-compare} following their descriptions, presented next.

\subsubsection{The {\sf Sparse-Centers} algorithm (for expanding graphs)}\label{subsubsec:sparse-centers}
This algorithm
is based on the following idea. Suppose we can partition the graph vertices into
$\sqrt{\eps n}$ disjoint parts where each part is connected. If we now take a spanning
tree over each part and select one edge between every two parts that have an edge between
them, then we obtain a connected subgraph with at most $(1+\eps)n$ edges.

\paragraph{The partition.}
The partition is defined based on $\sqrt{\eps n}$ special {\em center\/} vertices,
which are selected uniformly at random. Each vertex is assigned to the closest
center (breaking ties according to a fixed order over the centers), unless it is
further than some threshold $\tr$ from all centers, in which case it is a singleton in
the partition. This definition ensures that for each center the subgraph induced on the vertices assigned to that center is connected.

\paragraph{The local algorithm.}
Given an edge $(u,v)$, the local algorithm
finds the centers to which $u$ and $v$ are assigned (or determines that they are singletons).
This is simply done by performing a Breadth First Search (BFS) to distance at most $\tr$ from each of the two vertices, until
a center is encountered (or no center is encountered, in which case we get a singleton).
If one of the two vertices is a singleton, then $(u,v)$ is always taken to the spanning subgraph.
Otherwise, if $u$ and $v$ are assigned to the same center, $\z$, then the algorithm determines whether the
edge between them belongs to the BFS-tree rooted at $\z$ that spans the vertices assigned to $\z$.
(Since there may be many such trees, we consider
a fixed choice, based on the vertices' ids.)

To this end it is actually  not necessary to
determine for each vertex at distance at most $\tr$ from $\z$
whether it is assigned to $\z$ or not (which would be too costly in terms of the local algorithm's query complexity).
Rather, given that it is known that both $u$ and $v$ are assigned to $\z$, it suffices to compare the distances of $u$ and $v$ to $\z$,
and if they differ by $1$, to determine the distances between the neighbors of the further vertex, and $\z$.

If $u$ and $v$ belong to
different centers, $\z(u)$ and $\z(v)$, respectively, then the algorithm determines whether $(u,v)$ is the single edge
allowed to connect the parts corresponding to the centers $\z(u)$ and $\z(v)$. This is done according to a prespecified
rule which, once again, can be decided without  identifying exactly which vertices belong to
the two parts.
Specifically, consider the shortest path between $\z(u)$ and $\z(v)$ that is first in lexicographic order,
denoted $P(\z(u),\z(v))$. Observe that due to the edge $(u,v)$, this path has length at most $2t+1$.
Therefore, 
it can be found efficiently by performing a BFS to depth $\tr$ from both $\z(u)$ and $\z(v)$.
If the edge between $u$ and $v$ belongs to $P(\z(u),\z(v))$, then it is taken to the spanning graph, and otherwise it is not taken. Indeed it may be the case that no edge between a vertex assigned to $\z(u)$ and a vertex assigned to $\z(v)$ is included in the spanning graph, despite the existence of such edges. This occurs if $P(\z(u),\z(v))$ passes through vertices that are assigned to other centers.
As a consequence, the argument that the subgraph $G'$ is connected is required to be more subtle. Essentially, we show by induction that
$\z(u)$ and $\z(v)$ will be connected, but possibly by a path that passes through other centers.






\paragraph{The complexity of the algorithm.}
From the above description one can see that there is a certain tension between the complexity
of the algorithm and the number of edges in the spanning subgraph $G'$.
 On one hand, the
threshold $\tr$ should be such that with high probability (over the choice
of the centers) almost all vertices
have a center at distance at most $\tr$
from them. Thus we need a lower bound (of roughly $\sqrt{n/\eps}$) on the size of the distance-$\tr$
neighborhood of (most) vertices. On the other hand,  we also need an upper bound on the
size of such neighborhoods
so that we can efficiently determine
which edges are selected between parts.
Hence, this algorithm works for graphs in which the sizes of the aforementioned
local neighborhoods do not vary by too much and its complexity (in terms of
the dependence on $n$) is nearly $\tilde{O}(\sqrt{n})$. In particular this property holds for good
expander graphs.

\paragraph{The threshold $\tr$.}
If the threshold $\tr$ is not given to the algorithm, then as a first step, the algorithm approximates $\tr$.
 Namely, it finds a threshold $\tr'$ that gives similar guarantees as $\tr$, both in terms of the sparsity of the output subgraph and in terms of the query complexity of the algorithm.

\subsubsection{The {\sf Dense-Centers} algorithm (for minor-free graphs)}\label{subsubsec:dense-centers}
In what follows we assume that the graph has a bounded degree, $d$, which is provided to the algorithm.
We then explain how to remove this assumption.

Similarly to the {\sf Sparse-Centers} algorithm,
 the  algorithm for families of graphs with a fixed excluded minor
is based on a global partition that is determined by a random set of centers.
However, the number of centers in this algorithm is linear in $n$ (hence we call it the {\sf Dense-Centers} algorithm), and the definition of the partition is more involved.
Also similarly to what was described in Section~\ref{subsubsec:sparse-centers},
given such a partition, the edge set $E'$ of the spanning subgraph $G'$ is
defined by taking a spanning tree in
each part, and a single edge between every pair of parts that have
at least one edge between them (though the choice of the edge is different).
Here, the minor-freeness of the graph, together with a bound on the number of parts,
ensure that the number of pairs of parts that have
at least one edge between them is at most $\eps n$, and therefore
$|E'| \leq (1+\eps)n$.

We now provide a high level description of the partition 
and the corresponding local algorithm.



\paragraph{Properties of the partition.}
We define a partition that has each of the following four
properties (which for simplicity are not precisely quantified in this
introductory text): (1) the number of parts in
the partition is not too large (though it is linear in $n$);
(2) the size of each part is not too large (it is polynomial in $d$, $h$ and $1/\eps$);
(3) each part induces a connected subgraph;
(4) for every vertex $v$ it is possibly to efficiently find (that is, in time polynomial in $d$, $h$ and $1/\eps$), the
subset of vertices in the part that $v$ belongs to.

\paragraph{An initial centers-based partition.}
Initially, the partition is defined (as in the {\sf Sparse-Centers} algorithm) by a selection of {\em centers\/}, where
the centers are selected randomly (though not completely independently).
Each vertex is initially assigned to the closest selected center. For an appropriate setting
of the probability that a vertex is selected to be a center, 
with high probability, this initial partition has the first property.
Namely, the number of parts in the partition is not too large.
By the definition of the partition, each part is connected, so that
the partition has the third property as well.
However,
the partition is not expected to have the second property.
That is, some parts may be too large.
Even for a simple graph
such as the line, we expect to see parts of size logarithmic in $n$.
In addition, the same example shows that the
fourth property may not hold, i.e., there may be many vertices for which
it takes super-constant time to find the part that the vertex belongs to.
To deal with these two problems,
we give a procedure for refining the partition in two phases, as explained next.

\paragraph{Refining the partition, phase 1 (and a structural lemma).}
A first refinement  is obtained by the separation of vertices that in order to reach their assigned center
in a  BFS, must observe a number of vertices that is above a
certain predetermined threshold, $k$.
Each of these {\em remote\/} vertices becomes a singleton
subset in the partition. We prove that with probability $1-1/\Omega(n)$, the number of
these remote vertices is not too large, so that the first property (concerning the number of parts)
is maintained with high probability.

The probabilistic analysis builds on
a structural lemma, which may be of independent interest.
The lemma upper bounds, for any given vertex $v$, the number of vertices $u$ such that
$v$ can be reached from $u$ by performing a BFS until at most $k$ vertices are observed.
This number in turn upper bounds the size of each part in the partition after the
aforementioned refinement.
While this upper bound is not polynomial in $k$ and $d$, it suffices for the
purposes of our probabilistic analysis (and we also show that there is
no polynomial upper bound). Specifically, we use this bound in order to obtain a bound on the variance of the number of remote vertices.

In addition to the first property, the third property (connectivity of parts) is also
maintained in the resulting refined partition, and  the refinement
partially addresses the fourth property,
as remote vertices can quickly determine that they are far from all centers.
In addition, the new parts of the partition will be of size $1$ and thus will
not violate the second property.

\paragraph{Refining the partition, phase 2.}
After the 
the first phase of the refinement, there might be some large parts remaining
so that the second and fourth properties are not necessarily satisfied.
We further partition large parts into smaller (connected) parts by a
certain decomposition based on the BFS-tree of each large part.

\paragraph{The local algorithm.}
Given an edge $(u,v)\in E$, the main task of the local algorithm is to find the two parts to which the vertices belong in the final partition. Once these parts are found, the algorithm can easily decide whether the edge between $u$ and $v$ should belong to $E'$ or not (since each part is small, and every vertex has a bounded degree).
In order to find the part
that $u$ (similarly, $v$) belongs to, the local algorithm does the following.
First it performs a BFS until it finds a center, or it sees at least $k$ vertices (none of which is a center).
In the latter case, $u$ is a singleton (remote) vertex. In the former case, the algorithm has found the
center that $u$ is assigned to in the initial partition, which we denote by $\cent(u)$. The algorithm next performs a BFS starting from $\cent(u)$. For each vertex that it encounters, it checks whether this vertex is assigned to
$\cent(u)$ (in the initial partition). If the number of vertices that are found to be assigned to $\cent(u)$ is
above a certain threshold, then the decomposition of the part assigned to $\cent(u)$ needs to be emulated locally.
We show that this can be done recursively in an efficient manner.

\paragraph{The unbounded-degree case.}
We next describe how to remove the dependence on the given degree bound $d$ and obtain an algorithm that works for unbounded-degree minor-free graphs.
For a degree threshold $\wtd = \wtd(h,\eps)$, we say that a vertex $u$ is a {\em heavy\/}
vertex if the degree of $u$ is greater than $\wtd$, and otherwise it is a {\em light\/} vertex. We denote the set of heavy vertices
by $\heavy$, and the set of light vertices by $\light$.
The degree threshold $\wtd$ is set so that $|\heavy|$ is sufficiently small, and (since the graph is minor-free) the total number of edges
between vertices in $\heavy$ is $\eps n/4$. This already implies that we can take to the sparse spanning graph
all edges between vertices in $\heavy$.

Consider the graph $G[\light]$ induced by $\light$.
Observe that given a vertex $v$, we can determine with a single 
query whether it is heavy or light (simply by checking if it has a neighbor at index $\wtd$ or not).
Furthermore, we can run a BFS on $G[\light]$ starting from any light vertex, by performing at most $\wtd\cdot q$ queries, where $q$ is the number of light vertices reached in the BFS. Also note that $G[\light]$ is not necessarily connected.

Suppose we run a slight variant of the algorithm for bounded-degree graphs on
$G[\light]$ with the degree $d$ set to $\wtd$. The variant is that within each connected component of size
less than $k$ (where the parameter $k$ is as defined for the bounded-degree case), we simply construct a (BFS) spanning tree rooted at the vertex with
the smallest rank (id)
in the component (this can be easily implemented in a local manner). For larger connected components,
the bounded-degree algorithm is run as is.
We thus obtain a graph that spans each connected component of $G[\light]$, and contains at most $|\light|+\eps n/4$ edges with high probability.

It remains to decide which edges between heavy and light vertices should be taken to the spanning graph.
For each part $P$ 
of the partition computed on $G[\light]$ (where each small connected component constitutes a part), we can select a single
edge connecting the part to a heavy vertex (if such an edge exists).
We say in such a case that the vertices in $P$ are {\em assigned\/} to $u$, and refer to the set of
vertices assigned to $u$, together with $u$ as the {\em cluster\/}  of $u$, denoted ${\cal C}(u)$.

The remaining decisions are with respect to edges between vertices in different clusters: a light vertex $v$ in one cluster and a heavy vertex $u$ in another.
This is the challenging part in the extension of the algorithm to unbounded-degree graphs.
Ideally, we would like to select one edge between every two clusters that have an edge between them.
The difficulty is that due to the high degree of heavy vertices, clusters may be very large, so we cannot afford to determine for a given vertex who are all the vertices in its cluster.
On the other hand, if we take all edges between clusters, then  sparsity may not be maintained.\footnote{To illustrate this, consider the following (planar) graph. It has two vertices, $u_1$ and $u_2$ whose degree is $D=(n-2)/2$ each. Let us denote the  neighbors of $u_1$ by $v_{1,1},\dots,v_{1,D}$,
and the neighbors of $u_2$ by $v_{2,1},\dots,v_{2,D}$, where these two subsets of vertices are disjoint
and there is an edge between $v_{1,j}$ and $v_{2,j}$ for each $j = 1,\dots,D$.
Observe that for this graph, all connected components of $G[\light]$ are small (with 2 vertices).
For any assignment of the light vertices to the heavy ones, the number of edges between clusters is
$n/2-1$.} We address this difficulty by selecting only a small fraction of edges between adjacent clusters, or all of them if there aren't too many  (see more details in Section~\ref{subsubsec:HL}).

\subsubsection{A discussion of the similarities and differences between the algorithms}\label{subsubsec:alg-compare}
As noted previously, our two local algorithms share several common features (but differ in others).
We next emphasize both the similarities and the differences.
\begin{enumerate}
\item The first and most basic similarity is that they are both based on a partition of the vertices into
connected subsets. Furthermore, the partition is (either fully or partially)
determined by Voronoi cells of centers, where the centers are chosen randomly, and (relatively few) cells of single vertices.
\item The second similarity is that the sparse spanning subgraph is defined by the edges of  BFS trees, one for
each part of the partition (rooted at the centers), and (at most) one edge  between every pair of parts,
(or relatively few in the unbounded-degree variant of the {\sf Dense-Centers} algorithm).
\item The first difference is in the number of centers (and hence the number of parts). For the {\sf Sparse-Centers}
algorithm this number grows like $\sqrt{n}$, and for the {\sf Dense-Centers} algorithm it is linear in $n$.
The bound on the number of parts in the case of the {\sf Sparse-Centers} algorithm immediately gives a bound on
the sparsity of the resulting spanning subgraph, while in the case of the {\sf Dense-Centers} algorithm we rely on
the minor-freeness of the graph.
\item The second difference is in the way singletons are defined. For the {\sf Sparse-Centers} algorithm these are
vertices that in their  bounded-distance neighborhood there is no center, and for the {\sf Dense-Centers} algorithm, these
are vertices that in their bounded-size neighborhood there is no center (where the neighborhood is allowed to have relatively large diameter).
\item
The third difference is in the way the edges between parts are selected, and identified (locally), and the related need to refine the partition for the {\sf Dense-Centers} algorithm.
Specifically,  given a vertex $u$, the {\sf Dense-Centers} algorithm finds all the vertices that belong to $u$'s part. (Here we are not referring yet to the clusters, which are defined in the unbounded-degree case, but rather only to the parts defined in the bounded-degree case.)
It is then easy to decide for an edge $(u,v)$ where $u$ and $v$ belong to different parts, whether the edge $(u,v)$ belongs to the spanning subgraph. However, this
requires to refine the partition so as to obtain sufficiently small parts, while being able to construct a part
in a local manner. On the other hand, the {\sf Sparse-Centers} algorithm decides whether an edge $(u,v)$ belongs to the spanning subgraph by identifying the centers that $u$ and $v$ are assigned to, but without determining the vertices in the corresponding parts (and the spanning subgraph might not include an edge between some pairs of parts even though there are edges between them).
\item The fourth difference is that the {\sf Dense-Centers} algorithm extends to unbounded-degree (minor-free) graphs, by applying an additional edge-sampling technique.
\end{enumerate}
Given the above,  the main challenge in the design and analysis of the {\sf Sparse-Centers} algorithm is in maintaining connectivity in a sufficiently efficient manner by avoiding the need to construct complete parts of the partition.
The main challenge in the design and analysis of the {\sf  Dense-Centers} algorithm is in proving  that
sparsity is obtained with  high probability (at least $1-\Omega(1/n)$). This requires a careful analysis of the number of singletons (which relies on structural properties of the graph), and in bounding the number of edges between clusters in the unbounded-degree variant of the algorithm.

\subsection{Related work}\label{subsec:related}

The work that is most closely related to the current work is~\cite{LMRRS}.
The upper bound in~\cite{LMRRS} builds on a part of the extended abstract~\cite{LRR14}
(while the current work builds on a different part).
It was shown in~\cite{LMRRS} that for families of graph that are, roughly speaking, sufficiency non-expanding, one can provide an LSSG algorithm with query complexity that is independent of $n$ (however, super-exponential in $1/\eps$).
This is achieved by simulating a localized version of Kruskal's algorithm.
On the negative side, it was also shown in~\cite{LMRRS} that for graphs with expansion properties that are a little better, there is {\em no local algorithm\/} that inspects a number of 
edges that is independent of $n$.


 A couple of additional LSSG algorithms for families of minor-free graphs, which we do not include in this paper, are presented in~\cite{LRR14}.
The first algorithm is based on a reduction to the partition oracle in~\cite{LR15} and the second one is designed for weighted graphs.
For the unweighted case, the  result presented here for minor-free graphs
significantly improves on what was shown in~\cite{LRR14}. Specifically,
the algorithm presented in~\cite{LRR14} has complexity
$(d/\eps)^{\poly(h)\log(1/\eps)}$ (as well as higher stretch, which in particular depends polynomially on $d$).
Furthermore, the success probability of the algorithm presented in this work (that is, the probability
that $G'$ has at most $(1+\eps)n$ edges), is $1-1/\Omega(n)$, while it was only ensured to be
a high constant in the \cite{LRR14} algorithm.

Graph partitioning for graphs with an excluded minor has been studied extensively (see e.g.,~\cite{AST90, KR10}), and was found useful in constructing spanners and distance oracles (see e.g.,~\cite{KKS11}).
Compared to previous results, the main disadvantage of our partition is that the edge cut is not guaranteed to be small. However, its main benefit is that it is designed to be constructed locally in an efficient manner.

\subsubsection{Followup work on LSSG}
Recently, Lenzen and Levi~\cite{LL17} provided an LSSG algorithm for general bounded degree graphs with query complexity $\tilde{O}(n^{2/3}) \cdot\poly(1/\eps, d)$, where $d$ is the bound on the maximum degree.
Parter et al.~\cite{PRVY19} provide local algorithms for constructing spanners in graphs with unbounded degree, part of their techniques are inspired by the algorithms for LSSG on bounded degree graphs.

\subsubsection{Local algorithms for other graph problems}\label{subsec:local}
The model of {\em local computation algorithms}
as used in this work, was defined by Rubinfeld
et al.~\cite{RTVX} (see also \cite{ARVX12} and survey in~\cite{LM17}).
Such algorithms for maximal independent set, hypergraph coloring,
$k$-CNF, approximated maximum matching and approximated minimum vertex cover for bipartite graphs are given in \cite{RTVX,ARVX12,MRVX12,MV13,EMR14,LRY16,FMS15}.
This model generalizes other models that have
been studied in various contexts, including locally
decodable codes (e.g.,~\cite{STV99}),
local decompression \cite{DLRR13},
and local filters/reconstructors
\cite{ACC+08,SS10,B08,KPS08,JR11,CS06a}.
Local computation algorithms that give approximate solutions
for various optimization problems on graphs, including vertex cover,
maximal matching, and other packing and covering problems,  can also be
derived from sublinear time algorithms for parameter estimation
\cite{PR,MR,NO,HKNO09,YYI}.

Campagna et al.~\cite{CGR13} provide a local reconstructor for connectivity.
Namely, given a graph which is almost connected, their reconstructor provides oracle access to the adjacency matrix of a connected graph which is close to the original graph.
We emphasize that our model is different from theirs, in that they allow
the addition of new edges to the graph, whereas our algorithms must provide spanning subgraphs
whose edges are present in the original input graph.

\subsubsection{Distributed algorithms}\label{subsec:dist}
The term {\em local algorithms} 
was originally coined in the distributed
context~\cite{MNS95,NS95,Lin92}.
As observed by Parnas and Ron~\cite{PR}, local distributed algorithms can be used to obtain local computation algorithms as defined
in this work, by simply emulating the distributed algorithm on a sufficiently large subgraph
of the graph $G$. However, while the main complexity measure in the distributed setting is
the number of rounds (and messages may be required to be of bounded size),
our main
complexity measure is the number of queries performed on the graph $G$.
By this standard reduction, the bound on the number of queries (and hence running time)
depends on the size of the queried subgraph and may
grow exponentially with the number of rounds. Therefore, this reduction gives
meaningful results only when the number of rounds is significantly smaller than
the diameter of the graph.

Ram and Vicari~\cite{Rm2011} studied the problem of constructing sparse spanning subgraphs in the distributed {\sf LOCAL} model~\cite{Lin92} and provided an algorithm that runs in $\min\{D(G), O(\log n)\}$ number of rounds where $D(G)$ denotes the diameter of $G$.

 In the next paragraphs, unless stated otherwise, we refer to algorithms in the {\sf CONGEST} model (in which the size of the messages is bounded by $O(\log n)$)~\cite{P00}. 
The problem of computing a minimum weight spanning tree in this model is a central
one.

Kutten and Peleg~\cite{KP98} provided an algorithm that works in $O(\sqrt{n} \log^* n + D)$ rounds, where $D$ denotes the diameter of the graph. Their result is nearly optimal in terms of the complexity in $n$, as shown by Peleg and Rubinovich~\cite{PR00} who provided a lower bound of $\Omega(\sqrt{n}/\log n)$ rounds (when the length of the messages must be bounded).

Another problem studied in the distributed setting that is related to the one
studied in this paper, is finding a sparse spanner.
The requirement for spanners is much stronger since the distortion of the distance should be as minimal as possible.
Thus, to achieve this property, it is usually the case that the number of edges of the spanner is super-linear in $n$.
Spanners with only $O(n)$ edges are usually referred to as {\em ultra-sparse spanners}.
Pettie~\cite{Pet10} was the first to provide a distributed algorithm for finding ultra-sparse spanners without requiring messages of unbounded length or $O(D)$ rounds. The number of rounds of his algorithm is $\log^{1+o(1)}n$.
Recently, Elkin and Neiman~\cite{EN16} introduced an algorithm that produces a spanner of size $n(1+o(1))$, as long as the number of rounds is $\omega(\log n)$.
However, in both cases, the standard reduction of~\cite{PR} yields a local algorithm with a trivial linear bound on the query complexity.

\subsubsection{Parallel algorithms}\label{subsec:para}
The problems of computing a spanning tree and a minimum weight spanning tree were studied extensively in the parallel computing model (see, e.g.,~\cite{BC05}, and the references therein).
However, these parallel algorithms have time complexity that is at least logarithmic in $n$ and therefore do not yield an efficient algorithm in the local computation model.
See~\cite{RTVX,ARVX12} for further discussion on the relationship between the ability to construct local computation algorithms
and the parallel complexity of a problem.

\subsubsection{Local cluster algorithms}\label{subsec:clust}
Local algorithms for graph theoretic problems have also been given for
PageRank computations on the web graph~\cite{JW03,Ber06,SBC+06,ACL06,ABC+08}.
Local graph partitioning algorithms
have been presented in~\cite{ST04,ACL06,AP09,ZLM13,OZ13},
which find subsets of vertices whose internal connections are
significantly richer than their external connections in time that
depends on the size of the cluster that they output.
Even when the size of the cluster is guaranteed to be small, it is
not obvious how to use these algorithms in the local computation setting
where the cluster decompositions must be consistent among
queries to all vertices.

\subsubsection{Other related sublinear-time approximation algorithms for graphs}\label{subsec:sublin}
The problem of estimating the weight of a minimum weight spanning tree in sublinear time
was considered by Chazelle, Rubinfeld and Trevisan~\cite{CRT}.
They describe an algorithm whose running time depends on the approximation parameter,
the average degree and the range of the weights, but does not directly depend on the number
of nodes.

\subsubsection{Organization}
We start with some preliminaries in Section~\ref{sec:prel}.
In Section~\ref{sec:lb} we prove our lower bound (an alternative, direct proof can be found in Appendix~\ref{sec:lowerbound}). In Sections~\ref{sec:centers} and~\ref{sec:upper} we describe and analyze our two algorithms. Since we introduce many notations throughout the paper, for the aid of the reader we provide two tables with these notations in Appendix~\ref{sec:notation}.

\section{Preliminaries}\label{sec:prel}
Let $G = (V,E)$ be a graph over $n$ vertices.
Each vertex $v\in V$ has an id, $id(v)$, where there is a full order over the ids.
We assume we have query access to the incidence-lists representation of $G$.
Namely, for any vertex $v$ and index $i \geq 1$ 
it is possible to obtain
the $i^{\rm th}$ neighbor of $v$ by performing a query to the graph
(if $v$ has less than $i$ neighbors, then
a special symbol is returned).\footnote{Graphs are allowed to have self-loops and multiple edges, but for our problem
we may assume that there are no self-loops
and multiple-edges (since the answer on a self-loop can always be negative,
and the same is true for all but at most one among a set of parallel edges).}

We denote the distance between two vertices $u$ and $v$ in $G$ by $\d_G(u,v)$.
For vertex $v \in V$ and an integer $r$,
let $\Gamma_r(v,G)$ denote the set of vertices at distance at most
$r$ from $v$, referred to as the \emph{distance-$r$ neighborhood of $v$},
and let $C_r(v,G)$ denote the subgraph of $G$ induced
by $\Gamma_r(v,G)$.
Let $n_r(G) \eqdef \max_{v\in V} |\Gamma_r(v,G)|$.
When the graph $G$ is clear from the context, we shall use the shorthand
$\d(u,v)$, $\Gamma_r(v)$ and $C_r(v)$ for $\d_G(u,v)$, $\Gamma_r(v,G)$ and $C_r(v,G)$, respectively.
We let $N(v)$ denote the set of neighbors of $v$.

For a subset of vertices $X$, we let $G[X]$ denote the subgraph of $G$ induced by $X$.

The total order over the vertices induces a total order (ranking) $\rho$ over
the edges of the graph in the following straightforward manner:
$\rho((u,v)) < \rho((u',v))$ if and only if $\min\{u,v\} < \min\{u',v'\}$
or $\min\{u,v\} = \min\{u',v'\}$
and $\max\{u,v\} < \max\{u',v'\}$ (recall that $V = [n]$).
The total order over the vertices also induces an order over those vertices visited by
a Breadth First Search (BFS) starting from any given vertex $v$, and whenever we refer to
a BFS, we mean that it is performed according to this order.

Recall that a graph $H$ is called a {\em minor\/} of a graph $G$ if $H$ is isomorphic to a graph that can be obtained by zero or more edge contractions on a subgraph of $G$. A graph $G$ is {\em $H$-minor free} if $H$ is not a minor of $G$.

The next definition and theorem are used to measure the sparsity of minor-free graphs.
\BD
The {\sf arboricty} of a graph is the minimum number of forests into which its edges can be partitioned.
\ED
The next theorem follows from~\cite{Mad68} and~\cite{Nas64}.
\BT\label{thm:minor}
Let $H $ be a fixed graph over $h$ vertices and let $G = (V,E)$ be an $H$-minor free graph.
Then the arboricity of $G$ is at most $c(h)$, where $c(h) = O(h\log h)$. In particular, $|E| \leq c(h)\cdot |V|$.
\ET

\medskip
We shall also make use of the following theorem regarding $t$-wise independent random variables.
Recall that a discrete random variable $\chi = (\chi_1, \ldots, \chi_{m})\in [q]^{m}$
(where $[q] \eqdef \{1,\dots,q\}$) is {\em $t$-wise independent\/} if for every subset
$I = (i_1, \ldots, i_{s}) \subseteq [m]$ such that $s \leq t$ and every set of values $(a_1, \ldots, a_{s}) \in [q]^{s}$,
$\Pr\left[\bigwedge_{j\in [s]} (\chi_{i_j} = a_j)\right] =   
   {\textstyle\prod}_{j\in [s]}\Pr[\chi_{i_j} = a_j]$.

The following theorem is implied by the proof of Proposition 6.3 in~\cite{ABI86}.
\BT[Implied by~\cite{ABI86}]\label{thm:twise}
For every $1\leq t \leq m$, there exists an explicit construction of
a $t$-wise independent random variable $\chi = (\chi_1, \ldots, \chi_m)$ for the following two cases:
\BE
\item $(\chi_1, \ldots, \chi_m) \in \{0,1\}^m$ and for every $i\in [m]$, $\Pr[\chi_i = 1] = a_i/b_i$  where $\{a_i\}_{i=1}^m, \{b_i\}_{i=1}^m$ are integers and $b_i \leq m$ for every $i\in [m]$.
\item $(\chi_1, \ldots, \chi_m) \in [q]^m$ and the distribution of $\chi_i$, for every $i\in [m]$, is the uniform distribution over $[q]$,  where $q\geq m$ is any prime power.
\EE
The seed length (the number of independent random bits used in the construction) is $O(t \log m)$.  
Moreover, for any $1\leq i \leq m$, $X_i$ can be computed in  $O(t)$ operations in the word-RAM model.
\ET

\section{A Lower bound for general bounded-degree graphs}\label{sec:lb}
\BT\label{thm:lb}
Any LSSG algorithm must have query complexity $\Omega(\sqrt{n})$.
This result holds for graphs with a constant degree bound $d$ and for constant $0 \leq \eps \leq 2d/3$.
Furthermore, the result holds even if the algorithm is allowed a constant failure probability
(i.e., the number of edges in the subgraph $G'$ may be larger than $(1+\eps)n$ with small
constant probability).
\ET
The theorem, with a slightly lower upper bound on $\eps$, can be proved based on a reduction
from one-sided error testing of cycle-freeness. We provide the reduction below.
In the appendix we give a direct proof for Theorem~\ref{thm:lb}, where in fact, we prove that a similar statement holds even when $G'$ is allowed to be disconnected with some constant probability.
In both proofs of the lower bound the construction is of constant-degree graphs with strong expansion properties and hence the lower bound holds even when restricted to these families of graphs.

\paragraph{A Reduction from testing Cycle-freeness with one-sided error.}
In~\cite[Proposition 3.4]{GR02} Goldreich and Ron showed that any one-sided error algorithm for testing cycle-freeness in bounded-degree graphs must perform $\Omega(\sqrt{n})$  queries. Such a tester must accept every cycle-free graph with probability 1,
and must reject with probability at least $2/3$ every graph that is $\eps$-far from being cycle free.
A graph is said to be $\eps$-far from being cycle-free, if more than $\eps d n$ edges must be removed
in order to make it cycle-free (where $d$ is the degree bound). In other words, the graph contains
at least $(1+\eps d)n$ edges.
The lower bound in~\cite{GR02} holds even if the input graph is promised to be connected, and
for constant $\eps$ and $d$.

We observe that a lower bound of $\Omega(\sqrt{n})$ queries for the LSSG problem is implied by this lower bound
for testing cycle-freeness.
To verify this, consider the following reduction. Given an LSSG algorithm $\calA$, one can obtain a tester $\mathcal{B}$ for cycle-freeness as follows.
On input graph $G= (V, E)$, $\mathcal{B}$ samples uniformly at random $\Theta(1/\eps)$
pairs $(v,i)$ where $v\in V$ and $i \in [d]$. It queries $G$ to obtain each of the corresponding edges (where
some may not exist), and then queries $\calA$ on all sampled edges, with sparsity parameter set to $\eps d/2$.
The algorithm $\mathcal{B}$ accepts if and only if $\calA$ returns YES on all the edges in the sample.

It remains to analyze $\mathcal{B}$.
Completeness: If $G$ is cycle-free, then $\calA$ must return YES on all  edges in $E$, and hence
$\mathcal{B}$ accepts with probability $1$.
Soundness: if $G$ is $\eps$-far from being cycle-free, then $\calA$ must, with probability
$1- 1/\Omega(n)$
return NO on at least $(\eps d - \eps d/2) n = \eps d /2$ edges.
Let ${\cal E}$ denote the event that indeed $\calA$ returns NO on at least $\eps d /2$ edges of $G$ (recall that these edges are determined only by $G$ and the randomness of $\calA$).
Thus, given that ${\cal E}$ occurs (which happens with probability $1- 1/\Omega(n)$), each pair $(v,i)$ that $\mathcal{B}$ samples is a witness for rejection with probability at least $\epsilon$.
Hence, with probability $1 - (1-\eps)^{\Theta(1/\eps)}$, $\mathcal{B}$ samples one of these edges and rejects.
Thus, $\mathcal{B}$ rejects with high constant probability, as desired.

\section{Graphs with high expansion: The {\sf Sparse-Centers} Algorithm}\label{sec:centers}
In this section we describe an algorithm that
gives meaningful results for graphs in which, roughly speaking,
the local neighborhood of
almost all vertices expands in a similar rate and $d$ is either a constant or grows very moderately as a function of $n$.
In particular this includes
graphs with high expansion.
In fact we only require that the graph expands quickly for small sets:
A graph $G$ is an {\em $(s, h)$-vertex expander} if for all sets $S$ of size at most $s$, $N^+(S)$ is of size at least $h\cdot  |S|$,
where $N^+(S)$ denotes the set of vertices in $S$ and vertices adjacent to vertices in $S$.
Define $h_s(G)$ to be the maximum $h$ such that $G$ is an $(s, h)$-vertex expander (hence, $1 \leq h_s(G) \leq d$).

Recall that for a vertex $v \in V$ and an integer $r$,
$\Gamma_r(v,G)$ denotes the set of vertices at distance at most
$r$ from $v$,
and $n_r(G) \eqdef \max_{v\in V} |\Gamma_r(v,G)|$.
We shall make use of the following simple claim.

\BCM\label{clm:expand}
For every two integers $\tr$ and $s$, $n_{\tr}(G) \geq \min \{(h_s(G))^{\tr}, s\}$\;.
\ECM
\BPF
By definition, for every integer $s$, and graph $G$ is an $(s, h_s(G))$-vertex expander. Thus, for every vertex $v$ and for every integer $r$ for which  $|\Gamma_r(v,G)| \leq s$, it holds that $|\Gamma_{r+1}(v,G)| \geq h_s(G) \cdot|\Gamma_{r+1}(v,G)|$ and so the claim follows.
\EPF

\noindent\\
We shall prove the following theorem (where Algorithm~\ref{alg:local-centers}
appears in Section~\ref{sub:localC}).
\BT\label{thm:expander}
Algorithm~\ref{alg:local-centers} is an LSSG algorithm.
For any graph $G$ over $n$ vertices, and for every $\eps > 0$,
its query complexity is $O(d^2 \cdot s^{\log_{h_s(G)} d})$
for $s = s(n,\eps) \eqdef 2\sqrt{2n/\eps} \ln n$.
The number of random bits the algorithm uses is $O(\sqrt{\eps n} \log n)$,
and its running time is the same as its query complexity up to a factor of $O(\log n)$.
\ET
By Theorem~\ref{thm:expander}, for bounded degree graphs with high expansion we get query and running time complexity nearly $O(n^{1/2})$.
In particular, if $h_s(G) = \Omega(d)$ then the complexity is $d^2 \cdot n^{1/2 + O(1/\log d)}$.
In fact, even for $h_s(G) \geq d^{1/2+1/\log n}$ the complexity is $o(n)$.
We remark that in the construction of our lower bound of $\Omega(n^{1/2})$, that appears in Section~\ref{sec:lowerbound}, we construct a pair of families of $d$-regular random graphs.
In both families, the expansion (of small sets) is $\Omega(d)$,
implying that for these families the gap between our lower bound and upper bound is at most $n^{O(1/\log d)}$.

\medskip\noindent
Our algorithm, the {\sf Sparse-Centers} algorithm (which appears as Algorithm~\ref{alg:local-centers} in Section~\ref{sub:localC}), is based on a global algorithm, which is presented in Section~\ref{sub:globalC}.
The (local) {\sf Sparse-Centers} algorithm appears in Section~\ref{sub:localC} and it is analyzed in
the proof of Theorem~\ref{thm:centers}.

\subsection{The global algorithm}\label{sub:globalC}
For a given parameter $\tr$, the global algorithm first defines a global partition of
(part or all of) the graph vertices in the following randomized manner.

\BE
\item Select $\ell \eqdef \sqrt{\eps n/2}$ {\em centers\/} uniformly and independently at random from $V$,
and denote them $\z_1,\dots,\z_\ell$.
\item
Initially, all vertices are {\em unassigned\/}.
\item For $i = 0, \ldots, \tr$, for  $j = 1, \ldots, \ell$:\\
    Let $L^i_j$ denote the vertices in the $i^{\rm th}$ level of the BFS tree of $\z_j$ (where
    $L^0_j = \{\z_j\}$).
    Assign to $\z_j$ all vertices in $L^i_j$ that were not yet assigned to any  other $\z_{j'}$.
\EE
Let $S(\z_j)$ denote the set of vertices that are assigned to the center $\z_j$. By the
above construction, the subgraph induced by $S(\z_j)$ is connected.
To verify this, consider any vertex $\x\in S(\z_j)$ and observe that the parent of $\x$ in the BFS tree of $\z_j$ has to be in $S(\z_j)$ as well (otherwise we get a contradiction to the fact that $\x\in S(\z_j)$).
\medskip\noindent
The subgraph $G' = (V,E')$ is defined as follows.
\BE
\item For each center $\z$, let $E'(\z)$ denote the edges of a BFS-tree that spans
the subgraph induced by $S(\z)$ (where the BFS-tree is determined by the order over the
ids of the vertices in $S(\z)$). For each center $\z$, put in $E'$ all edges in $E'(\z)$.
\item For each vertex $w$ that does not belong to any $S(\z)$ for a center $\z$,
  put in $E'$ all edges incident to $w$.
\item For each pair of centers $\z$ and $\z'$, let $P(\z,\z')$
be the shortest path between $\z$ and $\z'$ that has minimum lexicographic order
among all shortest paths
(as determined by the ids of the vertices on the path).
If all vertices on
this path belong either to $S(\z)$ or to $S(\z')$, then add to $E'$
the single edge $(x,y) \in   P(\z,\z')$ such that $x \in S(\z)$ and $y \in S(\z')$,
where we denote this edge by $e(\z,\z')$.
\EE
In what follows we shall prove that $G'$ is connected and that for $\tr$ that is sufficiently large, $G'$ is sparse with high probability as well.
We begin by proving the latter claim.
To this end we define a parameter that determines the minimum distance needed for most vertices to see roughly $\sqrt{n}$ vertices.
More formally, define $\tr_{\eps, d}(G)$ to be the minimum distance $\tr$ ensuring that all but an
$\eps/(2d)$-fraction of the vertices have at least $s$ vertices in their $\tr$-neighborhood.
That is,
\BEQ
\tr_{\eps, d}(G) \eqdef
    \min_\tr\left\{\left|\left\{v:|\Gamma_{\tr}(v)| \geq s\right\}\right|\geq
     \left(1-\eps/(2d)\right)|V|\right\}\;.
\EEQ
\BL
For $\tr\geq \tr_{\eps, d}(G)$ it holds that $|E'| \leq (1+\eps)n$ with probability at least $1-1/n$ over the random choice of centers.
\EL
\BPF
Since for $j = 1,\dots,\ell$ the sets $E'(\z_j)$ are disjoint, we
have that $\left|\bigcup_{j=1}^\ell E'(\z_j) \right| < n$.
Since there is at most one edge $(x,y)$ added to $E'$ for each pair of centers $\z,\z'$ and the number of centers is
$\ell$ ($= \sqrt{\eps n/2}$),  the total number of these edges in $E'$ is at most $\eps n/2$.
Finally,
Let $V_{\tr,s}\subseteq V$ denote the subset of the vertices, $v$,
such that $|\Gamma_{\tr}(v)| \geq s$.
Since the centers are selected uniformly, independently at random,
for each $w \in V_{\tr,s}$
the probability that no vertex in  $\Gamma_\tr(w)$ is selected
to be a center, is at most $(1- s/n)^\ell  < 1/n^2$.
By taking a union bound over all vertices in $V_{\tr,s}$, with probability at least $1- 1/n$,
every $w\in V_{\tr,s}$ is assigned to some center $\z$. Since the number of vertices
in $V\setminus V_{\tr,s}$ is at most $\eps n/(2d)$ and each contributes at most $d$ edges to $E'$,
we get the desired upper bound on $|E'|$.
\EPF

\BL\label{lem:G-prime-connect}
$G'$ is connected.
\EL
\BPF
It suffices to prove that there is a path in $G'$ between every pair of
centers $\z$ and $\z'$. This suffices because for each vertex $w$ that is assigned to some center
$\z$, there is a path between $w$ and $\z$ (in the BFS-tree of $\z$), and for each vertex $w$
that is not assigned to any center, all edges incident to $w$ belong to $E'$.
The proof proceeds by induction on $\d(\z, \z')$ and the sum of the ids of $\z$ and $\z'$ as follows.

For the base case, consider a pair of centers $\z$ and $\z'$ for which $\d(\z, \z') = 1$.
In this case, the shortest path $P(\z, \z')$ consists of a single edge $(\z,\z')$ (where
$\z \in S(\z)$ and $\z'\in S(\z')$ by the definition of $S(\cdot)$), implying that $(\z, \z')\in E'$.

For the induction step, consider a pair of centers $\z$ and $\z'$ for which
$\d(\z, \z') > 1$, and assume by induction that the claim holds for every pair of centers $(\y,\y')$ such that
either $\d(\y,\y') < \d(\z,\z')$ or $\d(\y,\y') = \d(\z,\z')$ and $id(\y)+id(\y') < id(\z) + id(\z')$.
Similarly to base case, if the set of vertices in
$P(\z, \z')$ is contained entirely in $S(\z)\cup S(\z')$, then $\z$ and $\z'$ are connected by construction.
Namely, $P(\z, \z') = (\z,x_1,\dots,x_q,x'_s,\dots,x'_{q'},\z')$ where $x_1,\dots,x_q \in S(\z)$
and $x'_1,\dots,x'_{q'} \in S(\z')$. The edge $(x_q,x'_{q'})$ was added to $E'$ and there are paths in
the BFS-trees of $\z$ and $\z'$ between $\z$ and $x_q$ and between $\z'$ and $x'_{q'}$, respectively.

Otherwise ($P(\z, \z')$ is not contained entirely in $S(\z)\cup S(\z')$), we consider two cases.
\BE
\item There exists a vertex $x$ in
$P(\z, \z')$, and a center (different from $\z$ and $\z'$), $\y$, such that  $x\in S(\y)$.
Note that this must be the case when $\d(\z,\z') \leq 2\tr+1$.
This implies that
 either $\d(x,\y) < \d(x, \z')$ or that $\d(x,\y) = \d(x,\z')$ and $id(\y) < id(\z')$.
 Hence, either
 $$\d(\z,\y) \leq \d(\z,x)+\d(x,\y) < \d(\z,x)+ \d(x,\z') = \d(\z,\z')$$
 or  $\d(\z,\y) = \d(\z,\z')$ and $id(\z) + id(\y) < id(\z)+id(\z')$. In either case we can apply
 the induction hypothesis to obtain that $\z$ and $\y$ are connected.
 A symmetric argument gives us that $\z'$ and $\y$ are connected.
\item Otherwise, all the vertices on the path $P(\z, \z')$ that do not belong to
$S(\z) \cup S(\z')$ are vertices that are not assigned to any center. Since $E'$
contains all edges incident to such vertices, $\z$ and $\z'$ and connected in this case as
well.
\EE
The proof of Lemma~\ref{lem:G-prime-connect} is completed.
\EPF

\subsection{The local algorithm}\label{sub:localC}
\begin{algorithm}[ht!]
\caption{{\bf -- {\sf Sparse-Centers} Algorithm (with known $\tr$)}}
For a random choice of $\ell = \sqrt{\eps n/2}$ centers, $\z_1,\dots,\z_\ell$ in $V$ (which is
fixed for all queries), and for a given parameter $\tr$,
on query $(u, v)$:
\BE
\item Perform a BFS to depth $\tr$ in $G$ from $u$ and from $v$.
\item If
either $\Gamma_\tr(u)\cap \{\z_1,\dots,\z_\ell\}=\emptyset$ or
$\Gamma_\tr(v)\cap \{\z_1,\dots,\z_\ell\} = \emptyset$, then
 return YES.\label{step:2}
\item Otherwise, let $\z(u)$ be the center closest to $u$ and let $\z(v)$ be the center closest
to $v$ (if there is more than one such center, break ties according to the order
$\z_1,\dots,\z_\ell$).
\item If $\z(u)=\z(v)=\z$ then do the following:
  \BI
  \item  If $\d(u,\z) = \d(v,\z)$, then return NO.
  \item If $\d(v,\z) = \d(u,\z)+1$, then consider all neighbors $w$ of $v$ on a shortest path between $v$ and $\z$.
  If there exists such a neighbor $w$ for which $id(w) < id(u)$, then return NO, otherwise, return YES. (The case $\d(u,\z)=\d(v,\z)+1$ is handled analogously).
  \EI\label{step:4}
\item If $\z(u) \neq \z(v)$, then perform a BFS to depth $\tr$ from both of the centers $\z(u)$ and $\z(v)$. 
Query also all neighbors of vertices at distance $\tr$ from $\z(u)$ and $\z(v)$ (note that for every vertex at distance less than $\tr$ from one of these vertices, all its neighbors were queried in the course of the corresponding BFS).
Find the shortest path between $\z(u)$ and $\z(v)$ that has the smallest lexicographic order
and denote it by $P(\z(u),\z(v))$.
Finding this path can be implemented by considering the subgraph induced by all queried edges. This subgraph contains all shortest paths between $\z(u)$ and $\z(v)$, from which the one having smallest lexicographic order can be selected.
Return YES if both $u$ and $v$ belong to $P(\z(u),\z(v))$. Otherwise, return NO.\label{step:5}
\EE
\label{alg:local-centers}
\end{algorithm}

\BT\label{thm:centers}
Algorithm~\ref{alg:local-centers}, when run with
$\tr \geq \tr_{\eps,d}(G)$, is an LSSG algorithm.  The query complexity and the running time of the algorithm are
$O(d\cdot n_{\tr}(G))$ and $O(d\cdot n_{\tr}(G)\cdot \log n)$, respectively.
The random seed it uses is
of size $O(\sqrt{\eps n}\log n)$.
\ET
\BPF
We prove the theorem by showing that Algorithm~\ref{alg:local-centers} is a local emulation of the global algorithm that appears in Section~\ref{sub:globalC}.
Given $u$ and $v$, by performing a BFS to depth
$\tr$ from each of the two vertices,
Algorithm~\ref{alg:local-centers} either finds the centers $\z(u)$ and $\z(v)$ that $u$ and
$v$ are (respectively) assigned to (by the global algorithm, for the same selection of centers),
or for at least one of them it finds no center in its distance-$\tr$ neighborhood. In the latter
case, the edge $(u,v)$ belongs to $E'$, and  Algorithm~\ref{alg:local-centers}
 returns a positive answer, as
required. In the former case, there are two subcases.
\BE
\item If $u$ and $v$ are assigned to the same center,
that is, $\z(u)=\z(v)=\z$, then Algorithm~\ref{alg:local-centers} checks whether the edge $(u,v)$ is an
edge in the BFS-tree
of $\z$ (i.e., $(u,v) \in E'(\z)$). If $u$ and $v$ are on the same level of the tree (i.e.,
are at the same distance from $\z$), then Algorithm~\ref{alg:local-centers} returns a
negative answer, as required. If $v$ is one level further than $u$, then
Algorithm~\ref{alg:local-centers} checks whether $v$ has another neighbor, $w$, that is
also assigned to $\z$, is on the same level as $u$ and has a smaller id than $u$.
Namely, a neighbor of $v$ that is on a shortest path between $v$ and $\z$ and has a smaller id than $u$.
If this is the case, then the edge $(u,v)$ does not belong to the tree (but rather the
edge $(w,v)$) so that the algorithm returns a negative answer. If no such neighbor of $v$
exists, then the algorithm returns a positive answer (as required).
\item If $u$ and $v$ are assigned to different centers, that is, $\z(u) \neq \z(v)$,
then Algorithm~\ref{alg:local-centers} determines whether $(u,v) = e(\z(u),\z(v))$ exactly
as defined in the global algorithm: The algorithm finds $P(\z(u), \z(v))$ and returns a positive answer if and only if $(u,v)$ belongs to $P(\z(u), \z(v))$.
Notice that since $u\in S(\z(u))$ and $v\in S(\z(v))$, if $(u,v)$ belongs to $P(\z(u), \z(v))$ 
then all the vertices on $P(\z(u), \z(v))$ belong to either $S(\z(u))$ or $S(\z(v))$.
This is implied by the fact that for every center $\z$ and a vertex $w$ that is assigned to $\z$,  it holds that every vertex on a shortest path between $\z$ and $w$ is also assigned to $\z$.
\EE
The bound  on the query complexity of Algorithm~\ref{alg:local-centers} follows directly by inspection of the algorithm.
The running time of performing Steps~\ref{step:2}-~\ref{step:4} is upper bounded by the query complexity of performing the BFS, i.e., $O(d\cdot n_{\tr}(G))$, times the time sufficient for determining if a vertex is a center.
The latter can be done in $O(\log n)$ time. To see this we explain next how the algorithm access the set of centers.
In order to select $\ell$ random centers, the algorithm 
uses a random seed of length $O(\sqrt{\eps n}\log n)$, which is interpreted as a sequence of $O(\sqrt{\eps n})$ uniformly selected random centers (with replacement).
If we first sort this sequence of random centers in $O(\sqrt{\eps n}\log n)$ time, then we can decide for each vertex if it is a center in $O(\log n)$ time. Hence, the running time of of performing Steps~\ref{step:2}-~\ref{step:4} is bounded by the query complexity of the algorithm up to a factor of $O(\log n)$.
The running time of Step~\ref{step:5} is linear in the size of the subgraph induced on the BFS trees (from both centers) and therefore is linear in the query complexity. Overall, we obtain that the running time is $O(\log n)$ times the query complexity, as desired.
\EPF

\subsection{The parameter $\tr$}
Recall that Algorithm~\ref{alg:local-centers} is given a parameter $\tr$ that determines the depth of the BFS that the algorithm performs.
By Theorem~\ref{thm:centers} is suffices to require that $\tr \geq \tr_{\eps, d}(G)$ in order to ensure that the spanning graph obtained by the algorithm is sparse.
For the case that $\tr$ is not given in advance we describe next how to compute $\tr$ such that with probability at least $1-1/n$ it holds that
\BEQ\label{eq:approxk}
\tr_{\eps, d}(G) \leq \tr \leq \tr'_{\eps, d}(G)
    \;,
\EEQ
where $\tr'_{\eps, d}(G) = \min_\tr\left\{\left|\left\{v:\Gamma_\tr(v) \geq s\right\}\right|\geq
     \left(1-\eps/(4d)\right)|V|\right\}$.
Select uniformly at random $q= \Theta(\log n/\eps^2)$ vertices from $V$.
Let $x_1, \ldots, x_q$ denote the selected vertices.
For each vertex $x_j$ in the sample, let $\tr_j = \min_\tr\{\Gamma_\tr(x_j) \geq s\}$.
Assume without loss of generality that $\tr_1 \leq \ldots, \leq \tr_q$ and set $\tr = \tr_{\lceil1-\frac{3\eps}{8d}\rceil}$.
By Chernoff's inequality we obtain that with probability greater than $1-1/n$ Equation~(\ref{eq:approxk}) holds.

\medskip\noindent
We are now ready to prove Theorem~\ref{thm:expander}.

\BPFOF{Theorem~\ref{thm:expander}}
Assume without loss of generality that $\tr_{\eps,d}(G)$ is unknown and we run Algorithm~\ref{alg:local-centers} with $\tr$ such that $\tr_{\eps,d}(G) \leq \tr \leq \tr'_{\eps,d}(G)$.
Recall that $n_{\tr}(G)$ is the maximum size of a
distance-$\tr$
neighborhood in $G$ and that $h_s(G)$ is the maximum $h$ such that $G$ is an $(s, h)$-vertex expander.
By Claim~\ref{clm:expand}, $n_{\tr}(G) \geq \min \{(h_s(G))^{\tr}, s\}$. By the definition of $\tr'_{\eps, d}(G)$ we obtain that $(h_s(G))^{\tr'_{\eps, d}-1} \leq s$, since otherwise we get a contradiction to the minimality of  $\tr'_{\eps, d}$. Thus, we obtain that $(h_s(G))^{\tr-1}  \leq s$ and so $\tr \leq \frac{\log s}{\log h_s(G)} + 1$.
On the other hand, since the degree is bounded by $d$, it holds that $n_{\tr}(G) \leq 1+ d^{\tr}$.
By Theorem~\ref{thm:centers}, the query complexity and the running time of the algorithm are
$O(d\cdot n_{\tr}(G))$ and $O(d\cdot n_{\tr}(G)\cdot\log n)$, respectively. Hence, we get a bound of $O(d \cdot (1+ d^{\frac{\log s}{\log h_s(G)}+1}))$ and $O(d \cdot (1+ d^{\frac{\log s}{\log h_s(G)}+1})\log n)$ on the query complexity and the running time of the algorithm, as desired.
\EPFOF

\section{The algorithm for graphs with an excluded minor}\label{sec:upper}

In this section we prove the following theorem.
We say that a family of graphs is {\em minor-free\/} if the graphs in the family exclude some fixed minor.

\BT\label{thm:main1}
There exists an LSSG algorithm
for any family of minor-free graphs.
The query complexity and running time of the algorithm
are
$\poly(h/\eps)$,
where $h$ is the number of vertices in the excluded minor.
The algorithm uses $\poly(h/\eps)\cdot \log(n)$ random
bits, and
the stretch factor of the sparse spanning graph $G'$ is $\tilde{O}(h/\eps)$.
\ET
For the sake of simplicity, we did not try to optimize the various complexity measures of the LSSG algorithm (query complexity, running time, and number of random bits it uses). Hence, the
exponents in the polynomial expressions for these complexities are
quite high, 
and can probably be improved.

In order to prove Theorem~\ref{thm:main1}, we first design an LSSG algorithm for (minor-free) graphs whose maximum degree is upper bounded by a given integer $d$. Thereafter, we show how to adapt this algorithm to graphs with unbounded degree. We begin with proving the following theorem,
where Algorithm~\ref{alg:ssg} is provided in Section~\ref{subsec:dense-local}.

\BT\label{thm:main}
Algorithm~\ref{alg:ssg} is an LSSG algorithm
for the family of 
graphs that exclude a minor over $h$ vertices.
The query complexity of the algorithm is $\tilde{O}((h/\eps)^4 d^5)$ and its running time is
$\tilde{O}((h/\eps)^5 d^5)$.
The algorithm uses
$\tilde{O}((h/\eps)d\log n)$ random bits, and the stretch factor of the sparse spanning graph $G'$ is $\tilde{O}(h/\eps)$.
\ET

In Sections~\ref{subsec:partition}--\ref{subsec:dense-local} 
we describe our LSSG algorithm (Algorithm~\ref{alg:ssg}) for graphs with degree bounded by $d$.
In Section~\ref{subsec:unbounded}
we explain how to use this algorithm as a subroutine in an LSSG algorithm for graphs with unbounded degree.
For this purpose we require that Algorithm~\ref{alg:ssg} work also for graphs that are not connected.
Namely, the guarantee for the algorithm is that every connected component in $G$ remains connected in $G'$, and that $|E'| \leq (1+\eps)\cdot n$.
More precisely, since we are interested in bounding the total number of edges in the final
sparse spanning subgraph by $(1+\eps)\cdot n$, we shall actually ensure (with high probability) that
Algorithm~\ref{alg:ssg} answers positively on at most $(1+\eps/4)\cdot n$ edges.

We start by describing a global partition of the vertices. We later explain how to locally generate this partition and 
design our algorithm (and the sparse subgraph it defines), based on this partition.

\subsection{The partition $\calP$}\label{subsec:partition}
The partition described in this subsection, denoted by $\calP$, is a result of a random process.
We  describe how the partition is obtained in three steps where in each step we refine the partition from the previous step.
The partition is defined based on two parameters: $\gamma \in (0,1)$,
and an integer $k>1$, which are set subsequently (as a function of
$d$, $\eps$ and $h$).
Another central ingredient in the definition of the partition is a subset $\W\subseteq V$ of vertices. This subset is selected randomly as follows:
each vertex 
$v\in V$
draws a $\gamma$-biased bit, denoted $b_v$, and $v\in \W$ if an only if $b_v  = 1$. The joint-distribution of 
$(b_v)_{v\in V}$
is $t$-wise independent where $t \eqdef 2kd$.
(The reason that the choice of $\W$ is determined by a $t$-wise independent distribution rather than
an $n$-wise independent distribution is so as to bound the number of random bits used by the local emulation of
the global algorithm.)
In the next three subsections (\ref{subsubsec:1st}--\ref{subsubsec:3rd}) we explain
how the partition $\calP$ is defined, given the choice of $\W$.

\subsubsection{First step}\label{subsubsec:1st}
We begin with some notation.
For a vertex $v\in V$, let $\cc(v)$ denote the subset of vertices in the connected component of $v$ in $G$.
Let $Z$ denote the set of vertices $v$ such that $|\cc(v)| \geq k$.
Given 
a vertex $v\in V$, we define the {\em center\/} of $v$ with respect to $\W$, denoted
$\cent(v)$ as follows.
\BE
\item If $v \notin Z$, or if $v \in Z$ and $\cc(v) \cap W = \emptyset$, then $\cent(v)$ is  the vertex with the minimum
id (identifier) in $\cc(v)$.
\item 
If $v \in Z$ and $\cc(v) \cap W \neq \emptyset$, then $\cent(v)$ is
the vertex in $\W$ that is closest to $v$, breaking ties using vertex ids. That is,
$\cent(v)$ is the vertex with the minimum identifier
in the subset $\Big\{y\in \W : \d(y, v) = \min_{w\in \W} \{\d(w,v)\}\Big\}$.
\EE

For each $w\in \W$ we define the {\em cell\/} of $w$ with respect to $\W$ as
$\cell(w) \eqdef \{v\in V: \cent(v) = w \}$.
Namely, the set of vertices in $\cell(w)$ consists of the vertices that are closer to $w$ 
than to any other vertex in $\W$ (where ties are broken according to the ids of the vertices).
For every vertex $w \in V$ such that $w$ has the minimum identifier in $\cc(w)$, and $\cc(w) \cap W = \emptyset$, $\cell(w)$ is defined the same way (namely, as the subset of all vertices $v$ for which $\cent(v) = w$).
Notice that these cells form a partition of $V$.

\subsubsection{Second step}\label{subsubsec:2nd}
In this step we identify a subset of {\em special\/} vertices, which we call the {\em remote vertices\/} and make these vertices singletons in the partition $\calP$.
The set of remote vertices, $R$, which is a subset of $Z$, is defined with respect to $\W$ and the integer parameter $k$ as described next.
For every vertex $v \in Z$, let $\ell_k(v)$ be the minimum integer $\ell$ such that the BFS tree rooted at $v$ of depth $\ell$ has size at least $k$.
Let $B_k(v)$ be the set of vertices in the BFS tree rooted at $v$ of depth $\ell_k(v)$.
We define $R = \{v\in Z : B_k(v) \cap \W = \emptyset\}$, i.e., those vertices $v$ for which $B_k(v)$ does not contain a vertex (center) in $\W$.
Clearly, a vertex can identify efficiently if it is in $R$ by probing at most $kd$ vertices and checking whether they intersect $\W$.
In Section~\ref{subsec:b-remote} we obtain a bound on the size of $R$.

\subsubsection{Third step}\label{subsubsec:3rd}
In this step we decompose cells that are 
too big. We first argue that the cells are still connected (after the removal of the vertices in $R$ from all original cells defined in Section~\ref{subsubsec:1st}). Thereafter we will use a procedure of tree decomposition in order to break the cells into smaller parts.
\BCM\label{lem:shrt}
For every $w \in \W$, the subgraph induced by $\cell(w) \setminus R$ is connected.
Furthermore, for every $v \in \cell(w) \setminus R$, the subgraph induced by $\cell(w) \setminus R$ contains all vertices that belong to the shortest paths between $v$ and $w$.
\ECM
\BPF
Fix $w \in \W$, and consider any
$v\in \cell(w) \setminus R$. We will prove that the subgraph induced by $\cell(w) \setminus R$ contains all vertices on the  shortest paths between $v$ and $w$ and this will imply the connectivity as well.
The proof is by induction on the distance to $w$.
In the base case $v =  w$. In this case $\cell(w) \setminus R$ clearly contains a path between $v$ to itself because it contains $v$.
Otherwise, we shall show that for any $u\in N(v)$ for which $\d(u, w) < \d(v, w)$ it holds that $u \in \cell(w) \setminus R$.
The proof will then follow by induction.
Let $\Path$ be a shortest path between $v$ and $w$ and let $(v,u)\in E$ denote the first edge in $\Path$.
We first observe that $\cent(u) = w$ and thus $u\in \cell(w)$.
Assume otherwise and conclude that
there is a vertex in $w' \in \W$ for which either $\d(v,w') < \d(v,w)$ or
$\d(v ,w) = \d(v, w')$ and $id(w') < id(w)$, in contradiction to the fact that $\cent(v) = w$ (see the definition of $\cent(\cdot)$ in Section~\ref{subsubsec:1st}).
Since $u$ is on a shortest path between $v$ and $w$ it follows that
\BEQ
\Gamma_{\d(u, w)-1}(u) \subseteq \Gamma_{\d(v, w)-1}(v)\;.\label{eq:gamma}
\EEQ
From the fact that $v \notin R$ it follows that $|\Gamma_{\d(v, w)-1}(v)| \leq k$ and hence from Equation~\eqref{eq:gamma} it follows that $|\Gamma_{\d(u, w)-1}(u)| \leq k$ and so $u \notin R$ as well.
We conclude that $u \in \cell(w) \setminus R$ and $\d(u, w) = \d(v , w) - 1$ as desired.
\EPF

We shall use the following notation. For each $w\in \W$ let $\calT(w)$ denote the BFS tree rooted at $w$ of the subgraph induced by $\cell(w) \setminus R$
(recall that the BFS is performed by exploring the vertices according to the order of their identifiers).
To obtain the final refinement of our partition, for each $w\in \W$ such that $|\calT(w)| > \sz$, we run Algorithm~\ref{alg:treed} on $\calT(w)$, $w$ and $\sz$.

\begin{algorithm}
{\small
\caption{{\bf -- Recursive Tree decomposition}}\label{alg:treed}
\textbf{Input:} A tree $\calT$, the root of the tree $v$ and an integer $\sz$.\\
\textbf{Output:} A decomposition of $\calT$ into subtrees, where each subtree is assigned a (sub-)center.
\BE
\item Initialize the set of vertices of the current part $Q :=\emptyset$.
\item \label{step:bfs} Perform a BFS starting from $v$ and stop at level $\ell \eqdef \ell_\sz(v)$ (see the definition of $\ell_\sz(\cdot)$ in the Second Step). Add to $Q$ all the vertices explored in the BFS.
\item Let $S$ denote the set of all the children of the vertices in the $\ell^{\rm th}$ level of the BFS (namely, all the vertices in level $\ell+1$).
\item For each vertex $u \in S$:
\BE
\item \label{step:check}If the subtree rooted at $u$, $\calT_u$, has size at least $\sz$, then disconnect this subtree from $\calT$ and continue to decompose by recursing on input $\calT_u$, $u$ and $\sz$.
\item Otherwise, add the vertices of $\calT_u$ to $Q$.
\EE
\item \label{step:q} Set $v$ to be the {\em sub-center} of all the vertices in $Q$.
\EE
}
\end{algorithm}

\subsection{The edge set}\label{subsec:edges}
Given the partition $\calP$ 
defined in the previous subsection~(\ref{subsec:partition}),
we define the
edge set, $E'$, of our sparse spanning graph in the following simple manner.
In each part of $\calP$ that is not a singleton, we take a spanning tree. Specifically, we take the BFS-tree rooted at the sub-center of that part (see Algorithm~\ref{alg:treed}, Step~\ref{step:q}).
For every pair of parts of $X, Y \in \calP$, if the set
$E(X,Y) \eqdef (x,y)\in E : x\in X \text{ and } y \in Y\}$ is not empty, then we add to $E'$
the edge $e \in E(X,Y)$ with minimal ranking (where the ranking over edges is as defined in
Section~\ref{sec:prel}).
Clearly every connected component of $G$ is spanned by $E'$. We would like to bound the size of $E'$. To this end we will use
the following claim.
\BCM\label{clm:part-size}
Let $\calP'$ denote the partition $\calP$ when restricted to $Z$.
The number of parts in $\calP'$ is bounded as follows:
$|\calP'| \leq |\W| + |R| + \frac{n}{\sz}$.
\ECM
\BPF
Consider the three steps of refinement of the partition.
Clearly, after the first step the number of cells for which some $w\in W$ is the center is at most $|\W|$.
Therefore, after the second step, the size of the partition when restricted to $Z$, is at most $|\W| + |R|$. Finally, since in the last step we break only parts
whose size is greater than $\sz$ into 
smaller parts that are of size at least $\sz$, we obtain that the number of new parts that are introduced in this step is at most $n/\sz$. The claim follows.
\EPF

The next lemma establishes the connection between the size of $\calP'$ and the sparsity of $G'$.
\BCM\label{lem:Eprime-size}
For an input graph $G$ that is $H$-minor free for a graph $H$ over $h$ vertices,
$$|E'| 
   < n + |\calP'| \cdot c(h)\;,$$
where $c(h)$ is a bound on the arboricity of $G$ (as defined in 
   Theorem~\ref{thm:minor})
   and $\calP'$ is as defined in Claim~\ref{clm:part-size}.
\ECM
\BPF
Since for each $X\in \calP$ the subgraph induced by $X$ is connected, we can contract each part in $\calP$ and obtain an $H$-minor free graph.
The number of vertices in this graph is $|\calP|$. If we replace multi-edges with single edges, then by 
the bound on the number of edges of $H$-minor free graph (see Theorem~\ref{thm:minor}), we obtain that the number of edges in this graph is at most $ |\calP'| \cdot c(h)$ (observe that every part that is in $\calP$ but not in $\calP'$ becomes an isolated vertex in the contracted graph).
Finally, since the total number of edges in the union of spanning trees of each part is
$n-|\calP| < n$, we obtain the desired result.
\EPF

\subsection{Bounding the number of remote vertices}\label{subsec:b-remote}

In this subsection we prove the following lemma.
\BL\label{lem:R-size}
If $k = \Omega( (\log^2(1/\gamma) + \log d)/\gamma)$, then
with probability at least $1-\frac{1}{\Omega(n)}$ over the choice of $\W$, it holds that $|R| \leq \gamma n$.
\EL

We note that by Lemma~\ref{lem:R-size} it follows that there exists a distributed algorithm in the {\sf CONGEST} model for the LSSG problem with round complexity $O(k)$ which proceeds as follows. 
At the first step the set $\W$ is selected distributively. Afterwards, each vertex in $\W$ sends a unique token which is forwarded to neighbors by all vertices that did not receive a token so far. 
This goes on for $k$ steps. By the end of this process each vertex in the network knows to which part it belongs to (if it did not receive a token then it is in $R$ which means it is a singleton). 

On this note, we also remark that the partition-oracle of Hassidim et al.~\cite{HKNO09}, which can be used to obtain an LSSG algorithm, can be implemented in the {\sf LOCAL} model in time which is also $\poly(\eps^-1)$.  

In order to establish Lemma~\ref{lem:R-size}
we start by defining for every $v \in Z$,
\begin{equation}
Y_k(v) \eqdef \{u\in V: v \in B_k(u) \}\;.
\label{eq:Ykv-def}
\end{equation}
Informally, $Y_k(v)$ is the set of vertices that encounter $v$ while performing a BFS that stops after the first level in which the total number of explored vertices is at least $k$.

\medskip\noindent
We first establish the following simple claim.

\BCM\label{clm:Y-k}
For every vertex $u\in Y_k(v)$ and for every vertex $w$ that is on a shortest path
between $u$ and $v$, we have that $w \in Y_k(v)$.
\ECM
\BPF
Let $\ell = \d(u,v)$ and  $\ell' = \d(w,v)$, so that $\d(u,w) = \ell-\ell' \geq 1$.
Assume, contrary to the claim, that $w \notin Y_k(v)$. This implies that
$|\Gamma_{\ell'-1}(w)| \geq k$. But since $\Gamma_{\ell'-1}(w) \subseteq \Gamma_{\ell-1}(u)$,
we get that $|\Gamma_{\ell-1}(u)| \geq k$, contrary to the premise of the claim that
$u \in Y_k(v)$.
\EPF

\medskip
We now turn to upper bound the size of $Y_k(v)$.
\BL \label{lem:struct}
For every graph $G = (V,E)$ with degree bounded by $d$, and for every $v\in Z$,
$$
|Y_k(v)| \leq d^3 \cdot k^{\log k+1}\;.
$$
\EL
\BPF
Fix a vertex $v\in Z$.
For every $0 \leq j\leq k$, define $Y^j_k(v) \eqdef \{u\in Y_k(v):  \d(v, u) = j\}$.
Observe that $Y_k(v) = \bigcup_{j=0}^k Y^j_k(v)$.
Therefore, if we bound the size of $Y^j_k(v)$, for every $0 \leq j\leq k$, we will get a bound on the size of $Y_k(v)$.
Consider first any $3 \leq j < k$ and any vertex $u \in Y^j_k(v)$.
Recall that $\ell_k(u)$ is the minimum integer $\ell$ such that the BFS tree rooted at $v$ of depth $\ell$ has size at least $k$.
Since $j \leq \ell_k(u)$, it follows that $|\Gamma_{j-1}(u)| < k$.
Now consider a shortest path between $u$ and $v$ and let $w$ be the vertex on this path for which
$\d(u, w) = \lfloor(j-1)/2\rfloor$.
Denote $q \eqdef \d(w, v)$.
By 
Claim ~\ref{clm:Y-k}, $w\in Y_k(v)$, and by the definition of $q$,
$w \in Y^q_k(v)$.
Therefore,
\BEQ
|\Gamma_{q-1}(w)| \leq k\;.\label{eq:ell}
\EEQ
From the fact that $w$ is on the shortest path between $u$ and $v$ it also follows that
\BEQN
q &=& \d(v,u) - \d(u, w)
= j -  \lfloor(j-1)/2\rfloor \nonumber \\
&=& \lceil(j-1)/2 \rceil + 1  \label{eq:ceil} \\
&\geq& \lfloor(j-1)/2\rfloor + 1 \nonumber
= \d(u, w) + 1\;.
\EEQN
Therefore $q- 1 \geq \d(u,w)$ and so $u \in \Gamma_{q-1}(w)$.
It follows that
\BEQ
Y^j_k(v) \subseteq \bigcup_{w\in Y^q_k(v)} \Gamma_{q-1}(w)\;.\label{eq:yjv}
\EEQ
From Equations~\eqref{eq:ell} and~\eqref{eq:yjv} we get that $|Y^j_k(v)| \leq k \cdot |Y^q_k(v)|$.
For every $j \leq 3$ we have the trivial bound that $|Y^j_k(v)| \leq d^3$.
By combining with Equation~\eqref{eq:ceil} we get that $|Y^j_k(v)| \leq d^3 \cdot k^{\log j}$. Since $Y_k(v) = \bigcup_{j=0}^k Y^j_k(v)$ we obtain the desired bound.
\EPF

\medskip
While the bound on $|Y_k(v)|$ in Lemma~\ref{lem:struct} may not be tight, it suffices for
our purposes. One might conjecture that it is possible to prove a polynomial bound (in $k$ and $d$).
We show that this is not the case
(see Lemma~\ref{lem:Y-k-big} in the appendix).

\smallskip
We now use the bound in Lemma~\ref{lem:struct} in order to bound the number of remote vertices.

\smallskip
\BPFOF{Lemma~\ref{lem:R-size}}
Let $\chiW$ denote the characteristic vector of the set $\W$.
For a subset $S \subseteq V$, let $\chiW(S)$ denote the projection of $\chiW$ onto $S$.
That is, $\chiW(S)$ is a vector of length $|S|$ indicating for each $x\in S$ whether
$\chiW(x) = 1$.

For each vertex $v\in V$ define a random variable $\zeta_v$ indicating whether it is a remote vertex with respect to $\W$.
Recall that $v$ is remote if and only if $v\in Z$ and $B_k(v) \cap \W = \emptyset$.
Recalling that $\W$ is selected according to a $t$-wise independent distribution
where $t = 2kd$ and that $k \leq |B_k(v)| < k\cdot d$, we get that $\Pr[\zeta_v =1] \leq (1-\gamma)^k$.
We also define $S_v \eqdef \{u\in Z: B_k(u) \cap B_k(v) \neq \emptyset\}$.
Let $v\in V$. If $v\notin Z$ then the value of $\zeta_v$ is fixed and equals $0$.
Otherwise, observe that the value of $\zeta_v$ is determined by $\chiW(B_k(v))$.
Furthermore, since for every $v\in Z$ and $u \in Z\setminus S_v$, $\chiW(B_k(v))$ and $\chiW(B_k(u))$ are independent it follows that $\zeta_u$ and $\zeta_v$ are independent as well.
Hence, in this case $\cov[\zeta_v, \zeta_u] = 0$, and we obtain the following upper bound
on the variance of the number of remote vertices.
\BEQN
\Var\left[\sum_{v \in V} \zeta_v\right] &=& \sum_{(v, u)\in V} \cov[\zeta_v, \zeta_u]
\;=\;  \sum_{v\in V} \sum_{u\in S_v} \left(\Exp[\zeta_v\cdot \zeta_u] - \Exp[\zeta_u]\cdot\Exp[\zeta_v]\right)\nonumber\\
&\leq& \sum_{v\in Z} \sum_{u\in S_v} \Exp[\zeta_v \cdot \zeta_u |\zeta_v = 1]\cdot \Pr[\zeta_v = 1] 
\;\leq\; \sum_{v\in Z} |S_v| \cdot (1-\gamma)^k\;.\label{eq:sv}
\EEQN
By the definition of  $Y_k(\cdot)$ in Equation~(\ref{eq:Ykv-def}) it follows that
$S_v \subseteq \bigcup_{u\in B_k(v)} Y_k(u)$.
By Lemma~\ref{lem:struct}, $\max_{v\in V}\{|Y_k(v)|\} \leq d^3\cdot k^{\log k +1}$.
Therefore
\BEQ
|S_v| \leq |B_k(v)| \cdot d^3\cdot k^{\log k +1} \leq d^4\cdot k^{\log k +2}\;.\label{eq:bkv}
\EEQ

Hence, by Equations~\eqref{eq:sv} and~\eqref{eq:bkv} we get
$\Var\left[\sum_{v \in V} \zeta_v\right] \leq n d^4\cdot k^{\log k +2} \cdot (1-\gamma)^k$. Since $(1-\gamma)^k \leq e^{-\gamma k}$ we obtain that $\Var\left[\sum_{v \in V} \zeta_v\right] \leq \gamma^2 n$  for $k = \Omega( (\log^2(1/\gamma) + \log d)/\gamma)$.
Since for every $v\in V$, $\Pr\left[\zeta_v = 1\right] \leq (1-\gamma)^k \leq \gamma$, we get that $\Exp\left[\sum_{v \in V} \zeta_v\right] \leq \gamma n/2$.
By applying Chebyshev's inequality we get that
\BEQ
\Pr\left[\sum_{v \in V} \zeta_v \geq \Exp\left[\sum_{v \in V} \zeta_v\right] + \gamma n/2\right]\leq \frac{4\Var\left[\sum_{v \in V} \zeta_v\right]}{\gamma^2 n^2} \leq \frac{4}{n}\;.\nonumber
\EEQ
Since $|R| = \sum_{v \in V} \zeta_v$ it follows that $|R| < \gamma n$ with probability at least $1-(4/n)$, as desired.
\EPFOF

\subsection{The local algorithm}\label{subsec:dense-local}
In this subsection we provide Algorithm~\ref{alg:ssg}, which on input $e\in E$, locally decides whether $e\in E'$, 
as defined in Section~\ref{subsec:edges}, based on the (random, but not completely
independent) choice of $\W$. 
Given an edge $e=(u,v)$ we distinguish between two cases.
In the first case, $v\notin Z$. In this case, the center of $u$ is identical to the center of $v$ and can be found by performing a BFS from $v$ until the entire connected component $\cc(v)$ is revealed.
This is done in Step~\ref{step:firstfind} of Algorithm~\ref{alg:findc}.
In this case, $(u, v)$ belongs to $E'$ if and only if it belongs to the BFS tree rooted at $\cent(v)$.

In the second case, $v \in Z$.
In this case, the algorithm first finds, for each
$y \in \{u,v\}$, the center, $\cent(y)$, that $y$ is assigned to by the initial partition,
under the condition that $\cent(y) \in B_k(y)$. This is done in Step~\ref{step:secondfind} of Algorithm~\ref{alg:findc},
which simply performs a BFS starting from $y$ until it encounters a vertex in $\W$, or it reaches
level $\ell_k(y)$ without encountering such a vertex (in which case $y$ is a remote vertex).
Algorithm~\ref{alg:findc} assumes access to $\chiW$, which is implemented using the random seed
that defines the $t$-wise independent distribution, and hence determines $\W$.
If $y$ is not found to be a remote vertex, then Algorithm~\ref{alg:ssg} next determines to which sub-part
of $\cell(\cent(y))\setminus R$ does $y$ belong. This is done by emulating the tree decomposition of the
BFS tree rooted at $\cent(y)$ and induced by $\cell(\cent(y))\setminus R$. A central procedure
that is employed in this process is Algorithm~\ref{alg:bfs}. This algorithm is given
a vertex $v\in \W$, and a vertex $u$ in the BFS subtree rooted at $v$ and induced
by $\cell(v)\setminus R$. It returns the edges going from $u$ to its children in this tree,
by performing calls to Algorithm~\ref{alg:findc}.

\begin{algorithm}
{\small
\caption{{\bf -- {\sf Dense-Centers} Algorithm}}\label{alg:ssg}
\textbf{Input:} An edge $(u,v) \in E$. \\
\textbf{Output:} YES if $(u,v) \in E'$ and NO otherwise.
\BE
\item \label{step:find-part} For each $y\in \{u, v\}$ find the part that $y$ belongs to as follows:
\BE
\item Use Algorithm~\ref{alg:findc} to obtain $\cent(y)$.
\item If $\cent(y)$ is `null', then the part that $y$ belongs to is the singleton set $\{y\}$.
\item Otherwise, let $\calT$ denote the BFS tree rooted at $\cent(y)$ in the subgraph induced by $\cell(\cent(y)) \setminus R$. By Claim~\ref{lem:shrt} every shortest path between $\cent(y)$ and $v\in \cell(\cent(y)) \setminus R$ is contained in the subgraph induced on $\cell(\cent(y)) \setminus R$. Therefore the edges of $\calT$ can be explored via Algorithm~\ref{alg:bfs}.

   Reveal the part (the subset of vertices) that $y$ belongs to in $\calP$ (as defined in Section~\ref{subsec:partition}). Recall that the part of $y$ is the subtree of $\calT$ that contains $y$ after running Algorithm~\ref{alg:treed} on input $\calT$, $\cent(y)$ and $\sz$. This part can be revealed locally as follows.  \label{step:part}
\BE
\item Reveal the path between $\cent(y)$ and $y$ in $\calT$, denoted by $\Path(y)$. Since $\Path(y)$ is the shortest path between $y$ and $\cent(y)$ with the lexicographically smallest order, it can be revealed by performing a BFS from $y$ until $\cent(y)$ is encountered.
\item Run Algorithm~\ref{alg:treed} on $\calT$ while recursing in Step~\ref{step:check} only on the subtrees in which the root is contained in $\Path(y)$.
\EE
\EE
\item If $u$ and $v$ are in the same part,
then return YES iff the edge $(u, v)$ belongs to the BFS tree of that part.
\item Otherwise, return YES iff the edge $(u, v)$ is the 
edge with minimum rank connecting the two parts.
\EE
}
\end{algorithm}

\begin{algorithm}
{\small
\caption{{\bf -- Find Center}}\label{alg:findc}
\textbf{Input:} A vertex $v$ and an integer $k$. Query access to $\chiW$.\\
\textbf{Output:} $\cent(v)$ if $\cent(v) \in B_k(v)$ or $|\cc(v)| < k$, otherwise return `null'.
\BE
\item Perform a BFS from $v$ until at most $k$ vertices are revealed. \label{step:firstfind}
If the entire connected component of $v$ is revealed, then return the vertex with the minimum id from $\cc(v)$.
\item Otherwise, perform a BFS from $v$ until
the first level that contains a vertex in $\W$ or
until at least $k$ vertices are reached. That is, defining $\W_j \eqdef \Gamma_j(v) \cap \W$, stop at
level $\ell$ where
$\ell \eqdef \min\{ \ell_k(v), \min_{j}\{{\W_j \neq \emptyset}\}  \}$. \label{step:secondfind}
\item If  $\W_\ell = \emptyset$ then return  \textbf{`null'} ($v$ is remote).
\item Otherwise, return the vertex with the minimum id from $\W_\ell$.
\EE
}
\end{algorithm}

\medskip
\BPFOF{Theorem~\ref{thm:main}}
Let $c(h)$ as defined in Theorem~\ref{thm:minor} (namely, a bound on the arboricity of $H$-minor free graphs).
We set 
$\gamma = \eps/(16c(h))$ and
$k = c'\cdot(\log^2(1/\gamma) + \log d)/\gamma$, where $c'$ is a sufficiently large constant.
We start by bounding the size of $\W$ (with high probability).
Recall that $\W$ is selected according to $t$-wise independent
random variable over $\bitset^n$ where $\Pr[\chiW(i)=1] = \gamma$ for each $i\in [n]$
(and recall that $t = 2kd$).
By the definition of $\W$ we have that
$\Exp[|\W|] = \Exp\left[ \sum_{i\in [n]} \chiW(i)\right] = \gamma n$.
Since for every $1 \leq i <  j \leq n$, $\chiW(i)$ and $\chiW(j)$ are pairwise-independent,
we obtain that
$$\Var\left[\sum_{i\in [n]} \chiW(i)\right] = \sum_{i\in [n]} \Var \left[\chiW(i)\right] = \sum_{i\in [n]} \left(\Exp \left[\chi^2_{\mbox{\tiny\it W}}(i)\right] - \Exp \left[\chiW(i)\right]^2\right) = n\gamma(1-\gamma)\;.$$
Therefore,
by Chebyshev's inequality
$\Pr \left[|\W| \geq 2\gamma n\right] \leq \frac{1-\gamma}{\gamma n}$.
By Lemma~\ref{lem:R-size} (and the setting of $k$), with probability $1-1/\Omega(n)$,
$|R| \leq \gamma n$.
By Claims~\ref{clm:part-size} and~\ref{lem:Eprime-size} and the settings of
 $\gamma$ and $\sz$, we get that
 $$|E'| \leq n + (|W| + |R| + n/\sz)\cdot c(h) \leq n + 4\gamma\cdot c(h) \leq (1+\eps/4)n$$
 with probability $1-1/\Omega(n)$.

The connectivity in $G'$ of each connected component of $G$ follows by construction of $E'$. The claim about the stretch of $G'$ follows from the fact that the diameter of every part of $\calP$ is bounded by $2k$.

The number of vertices that Algorithm~\ref{alg:findc} 
traverses while performing a BFS (for any vertex $v$ it is called with) is at most $(k-1)d < kd$.
To verify this, observe that the algorithm continues to explore another level of the BFS only if strictly less than $k$ vertices were 
encountered so far.
Since the degree of each vertex is bounded by $d$, the query complexity of performing the BFS is bounded by $kd^2$.
Algorithm~\ref{alg:findc} makes at most $O(kd)$ accesses to $\chiW$. By Theorem~\ref{thm:twise}, each access to $\chiW$ takes $O(kd)$ time (as the distribution of $\chiW$ is $O(kd)$-wise independent). Hence, the running time of the algorithm is bounded by $O(k^2d^2)$.
The query complexity of Algorithm~\ref{alg:bfs} is upper bounded by the total query complexity
of at most $d$ calls to  Algorithm~\ref{alg:findc}, plus $d$ executions of a BFS until at most
$kd$ vertices are encountered (Steps~\ref{st:u-up} and~\ref{st:up-w}, which for simplicity
we account for separately from Algorithm~\ref{alg:findc}).
A similar bound holds for the running time.
Hence, the total query complexity and running time  of Algorithm~\ref{alg:bfs} are  $O(k d^3)$ and $O(k^2 d^3)$, respectively.

The size of any subtree returned by Algorithm~\ref{alg:treed} is upper bounded by $\sz^2  d^2$.
To verify this, recall that at the end of Step~\ref{step:bfs} of Algorithm~\ref{alg:treed}, at most $\sz d$ vertices were explored. Hence, the number of vertices that are incident to the explored vertices is at most  $\sz d^2$.
Thus, due to Step~\ref{step:check}, the total number of vertices in each part is at most $\sz^2 d^2$.
Recall that the distance of every vertex in $\calT$ from the root $v$ is at most $\sz$.
Hence, Step~\ref{step:check} can be implemented locally by calling Algorithm~\ref{alg:bfs} at most $\sz$ times. Thus, we obtain that the total query complexity and running time of
constructing each part is at most $O(k^3 d^5)$ and $O(k^4 d^5)$, respectively.
The query complexity and running time of Algorithm~\ref{alg:ssg} are dominated by those of Step~\ref{step:part}.
Observe that in Step~\ref{step:part} at most $|\Path(y)|$ parts are constructed.
Since $|\Path(y)|  \leq k$, we obtain that  the overall query complexity and running time  of Algorithm~\ref{alg:ssg} are bounded by $O(k^4 d^5)$ and $O(k^5 d^5)$, respectively.
By the setting of $k$ we obtain the final result.
\EPFOF

\begin{algorithm}
{\small
\caption{{\bf - Get BFS outgoing-edges endpoints}}\label{alg:bfs}
\textbf{Input:} $v\in \W$ and a vertex $u \in \calT(v)$ where $\calT(v)$ denotes the BFS tree induced by $\cell(v) \setminus R$ and rooted at $v$.\\
\textbf{Output:} The outgoing edges from $u$ in $\calT(v)$ (the orientation of the edges is from the root to the leaves).
\BE
\item Initialize $S: = \{v\}$  
\item For each $u' \in N(u)$, if 
 the following three conditions hold, then add $u'$ to $S$:
\BE
\item Algorithm~\ref{alg:findc} on input $u'$ returns $v$. Namely, $\cent(u') = v$.
\item \label{st:u-up} $u$ is on a shortest path between $u'$ and $v$. Namely, $\d(u', v)= \d(u, v) + 1$.
\item \label{st:up-w}
$u$ is the vertex with the minimum id among all vertices in  $\{w \in N(u'): \d(w, v) = \d(u', v) - 1\}$.
\EE
\item Return $S$.
\EE
}
\end{algorithm}


\subsection{Removing the dependence on the maximum degree}\label{subsec:unbounded}
In this subsection we show how to adapt Algorithm~\ref{alg:ssg} to work with unbounded degrees.
Let $c(h)$ be a bound on the arboricity of $H$-minor free graphs (as defined in Theorem~\ref{thm:minor}), where $h$ denotes the number of vertices in the excluded minor $H$.
We say that a vertex is {\em heavy} if its degree is greater than
$\wtd \eqdef 8(c(h))^2/\eps$.
Let $\heavy$ denote the set of {\em heavy} vertices and let $\light \eqdef V\setminus \heavy$ denote the set of {\em light} vertices.
As we explain next, we deal separately (and differently) with edges in $\heavy \times \heavy$, $\light \times \light$, and $\heavy \times \light$.

\subsubsection{The edges between pairs of heavy vertices}\label{subsubsec:HH}
We add all the edges in $G[\heavy]$, namely, all the edges between pairs of heavy vertices, to the spanning graph.
Since $G[\heavy]$ excludes $H$ as a minor, by Theorem~\ref{thm:minor},
the number of edges in $G[\heavy]$ is at most $c(h)\cdot |\heavy|$.
On the other hand, the size of $\heavy$ is at most
$\eps n/(4c(h))$ 
(otherwise, we obtain a contradiction to the size of $E$).
Hence, $G[\heavy]$ has at most $(\eps/4)n$ 
edges.

\subsubsection{The edges between pairs of light vertices}\label{subsubsec:LL}
Consider $G[\light]$, namely, the subgraph that remains after removing the vertices in $\heavy$ (and the edges incident to them).
By its definition, the maximum degree in $G[\light]$ is at most $\wtd$.
For an edge $(u, v)$ such that $u,v\in\light$, consider emulating the execution of Algorithm~\ref{alg:ssg} on $G[\light]$ given the query $(u,v)$. Observe that such an emulation can be easily implemented on $G$ at a multiplicative cost of at most $\wtd$.
The edge $(u,v)$ belongs to the spanning graph if and only if the emulation of Algorithm~\ref{alg:ssg} returns YES on query $(u, v)$.

\subsubsection{The edges between heavy and light vertices}\label{subsubsec:HL}
For every light vertex  $v$, let $\calP(v)$ denote the part that $v$ belongs to in the partition
$\calP$ of $G[\light]$ as computed in Step~\ref{step:part} of Algorithm~\ref{alg:ssg}.
If there is at least one edge $(u,v')$ between a heavy vertex $u$ and a (light) vertex $v' \in \calP(v)$, then
we {\em assign} $\calP(v)$ to the adjacent heavy vertex with the 
minimum id.
For a heavy vertex $u$, the {\em cluster} of $u$, denoted by $\calC(u)$, includes $u$ and all the vertices in the parts that are assigned to $u$.
We refer to $u$ as the {\em leader} of the vertices in $\calC(u)$.

On query $(u ,v)$ where $u$ is heavy and $v$ is light,
the query is answered as described in
Algorithm~\ref{alg:hle}.
We next discuss how the algorithm works, and give some high-level intuition for its correctness
(a formal proof appears in Section~\ref{subsubsec:sparsity}).

If $v$ belongs to the cluster of $u$, i.e., $v\in \calC(u)$, then
the answer to the query $(u,v)$ is YES if and only if $v$ has the minimum id among the vertices
in $\calP(v)$ that are adjacent to $u$.
Otherwise ($v$ does not belong to $\calC(u)$), let $w$ denote the leader of $v$  (so that $w\neq u$).
Consider all the edges in $\heavy \times \light$ between $\calC(u)$ and $\calC(w)$, whose set we denote by
$\tE(u, w)$, so that $(u,v) \in \tE(u, w)$. Observe that all  edges in $\tE(u, w)$ are
incident to either $u$ or $w$. 
We would have liked to include in $E'$ only a single edge from $\tE(u, w)$, say the edge with minimum rank
(recall that we consider a ranking over the edges based on the ids of their endpoints).
However, $u$ and $w$ may have a  high degree, so that we cannot afford querying all their incident edges in order to find the one with minimum rank in $\tE(u, w)$ (and answer the query on $(u,v)$ accordingly).

In order to save in the number of queries, we allow for more than a single edge in
$\tE(u, w)$ to be included in $E'$. Suppose we had access to uniformly selected
edges in $\tE(u, w)$ (at unit cost per edge). Then we could take a sample of such edges
and answer YES on $(u,v)$ if and only if no edge in the sample has rank smaller than $(u,v)$.
Roughly speaking, this would ensure
(with high probability) that only edges in $\tE(u, w)$ with relatively low rank will be
included in $E'$.

Since we do not have direct access to uniformly selected
edges in $\tE(u, w)$ we do the following. We take a random sample of edges incident to $u$ and a random sample of edges incident to $w$. For each selected edge $(u,y)$ (and analogously for selected edges $(w,y)$), we check what vertex is the leader of $y$. %
If the leader of $y$ is $w$, then we obtained an edge $(u,y)$ in
$\tE(u, w)$.
If the leader of $y$ is $u$, then we find all vertices in $\calP(y)$, and if some $z\in \calP(y)$
is a neighbor of $w$ (recall that $z$ has low degree), then we obtained an edge $(z,w)$ in $\tE(u, w)$.
(If the leader of $y$ is neither $w$ nor $u$, then the sampled edge is not useful for us.)
If we obtained an edge in  $\tE(u, w)$ that
has  rank smaller than $(u,v)$, then we  answer NO on $(u,v)$, otherwise we answer YES.

Conditioned on an edge in $\tE(u, w)$ being obtained, while the distribution on these edges is not uniform, we can lower bound the probability that each edge is selected, which suffices for our purposes.
If the probability that we obtain an edge in $\tE(u, w)$ is sufficiently large, then we are done
(as we can upper bound the total fraction of edges in $\tE(u, w)$ that are included in $E'$).
Otherwise, we may include in $E'$ all edges in $\tE(u, w)$,
where we upper bound the total number of such edges by
using the fact that $G$ is $H$-minor free.
For further details, see the
proof of Theorem~\ref{thm:main1}.

For each queried edge $(u,v)$ (where $u$ is heavy, $v$ is light, and
$v \in \calC(w)$ for $w\neq u$), the sampled edges incident to $u$ and $w$ are distributed uniformly
(among the edges incident to $u$ and $w$, respectively).
However, similarly to the choice of the subset $\W$ in the definition of the partition $\calP$ (see Section~\ref{subsec:partition}), in order to save in randomness,
the selection of edges that is performed for different queries is not done independently.
Rather, the samples are all determined by a random seed of $\poly(h/\eps)\log(n)$ bits.

\begin{algorithm}
{\small
\caption{{\bf - Heavy-Light Edges}}\label{alg:hle}
\textbf{Input:} query $(u ,v)$ where $u$ is heavy and $v$ is light.\\
\textbf{Output:} YES if $(u ,v) \in E'$ and NO otherwise.
\BE
\item Find the part of $v$, $\calP(v)$ (as described in Step~\ref{step:find-part} of Algorithm~\ref{alg:ssg}).
Find the leader of
$v$, denoted $w$ (by querying all edges incident to vertices in $\calP(v)$).
\item If $u = w$, then return YES if and only if
$v$ has minimum id among all vertices in $\calP(v)$ that are adjacent to $u$.
\item Sample a set $X_u$ of $q = \tilde{\Theta}((c^3(h) \sz^2 \wtd^3)/\eps^3)$ vertices that are adjacent to $u$,
where $\sz$ is as defined in the proof of Theorem~\ref{thm:main}, and similarly sample a set $X_w$ of $q$ vertices that are adjacent to $w$. \label{step:sam}
\item For each light vertex $y$ in $X_u$:
\BE
\item If its leader is $w$ and $\rho(u, y) < \rho(u, v)$, then return NO.
\item If its leader is $u$ and there is an edge $(z, w)$ such that
$\rho(z, w) < \rho(u, v)$ and $z \in \calP(y)$, then return NO.
\EE
\item For each light vertex $y$ in $X_w$:
\BE
\item If its leader is $u$ and $\rho(w, y) < \rho(u, v)$, then return NO.
\item If its leader is $w$ and there is an edge $(z, u)$ such that $\rho(z, u) < \rho(u, v)$ and $z \in \calP(y)$, then return NO.
\EE
\item 
        Return YES.
\EE
}
\end{algorithm}

\subsubsection{Correctness of the algorithm - Proof of Theorem~\ref{thm:main1}}\label{subsubsec:sparsity}
The connectivity of $G'$ follows directly from the construction of $E'$.
Namely: (1) Pairs of vertices that belong to the same connected components of $G[\light]$ are connected in $G'$
by the correctness of Algorithm~\ref{alg:ssg}.
(2) Pairs of vertices that
belong to the same cluster are connected either within their part in the partition
$\calP$ of $G[\light]$ or via their leader in the cluster.
(3) Pairs of vertices in different clusters (that are not in the same connected components of $G[\light]$)
are connected since there is at least one edge in $E'$ between every pair of clusters for which such an
edge exists.

By construction,
the stretch factor is at most a constant factor greater than the stretch factor of Algorithm~\ref{alg:ssg}.
To verify this, observe that the diameter of each cluster is bounded by $2k+2$ where $k$ is a bound on the stretch factor of Algorithm~\ref{alg:ssg}. Since for every pair of adjacent clusters there exists at least one edge in $G'$ which connects these clusters, we obtain that for each edge we remove from $(u, v) \in \heavy \times \light$ there exists in $G'$ a path of length at most $2k+2$ from $u$ to $v$.

We now turn to bound the total number of edges in $E'$. Let $F$ denote the set of edges in $E'\cap (\heavy \times \light)$ that are incident to two different clusters (namely, each endpoint belongs to a different cluster).
\BCM\label{clm:Ep-F}
With probability $1-1/\Omega(1/n)$,  $|E' \setminus F| \leq (1+(\eps/2))n$\;.
\ECM
\BPF
For a subset of vertices $U$, let $E'(U)$ denote the subset of edges in $E'$ between
vertices in $U$.
For each part $X$ in the partition $\calP$ of $G[\light]$ we add to $E'$ a spanning tree and so
we have that $|E'(X)| = |X|-1$. For a heavy vertex $u$, let $X_1,\dots,X_s$
be the parts assigned to it (so that $\calC(u) = \{u\} \cup \bigcup_{j=1}^s X_j$).
Since each part connects to its leader by a single edge (in $G'$), it follows that
$\left|\bigcup_{j=1}^s E'(\{u\}\cup X_j)\right| = |\calC(u)| - 1$.
As for the rest of the edges  in $(\light \times \light) \cap E'$,
by the proof of Theorem~\ref{thm:main}, with probability $1-1/\Omega(n)$ (over the choice of $\W$), their number is upper bounded by $(\eps/4)n$.
Since the number of edges in $\heavy \times \heavy$ is at most $(\eps/4) n$ (see details in Section~\ref{subsubsec:HH}), the claim follows.
\EPF

\BCM\label{clm:F}
With probability $1-1/\Omega(n)$, $|F| \leq (\eps/2)n$\;.
\ECM
\BPF
For each cluster $B$, we charge to $B$  a subset of the edges incident to $B$
so that the union of all the charged edges (over all clusters) contains $F$.
Our goal is to show that 
with probability $1-1/\Omega(n)$,
the total number of charged edges is $(\eps/2)n$.

Consider the  auxiliary graph, denoted $\wtG$, that results from contracting each cluster $B$ in $G$
into a  {\em mega-vertex\/} in $\wtG$, which we denote by $v(B)$.
For each pair of clusters $B$ and $B'$ such that $E(B,B')$ is non-empty, there is an edge $(v(B),v(B'))$ in  $\wtG$, which we refer to as a {\em mega edge\/}, and whose weight is $|E(B,B')|$. Since $G$ is $H$-minor free, so is $\wtG$.
By Theorem~\ref{thm:minor}, which bounds the arboricity of $H$-minor free graphs, we can partition the mega-edges of $\wtG$ into $c(h)$ forests.
Consider orienting the mega-edges of $\wtG$ according to this partition (from children to parents in the trees of these forests), so that each mega-vertex has at most $c(h)$ outgoing mega-edges.
For cluster $B$ and a cluster $B'$ such that $(v(B),v(B'))$ is an edge in $\wtG$ that is oriented from $v(B)$ to $v(B')$,
we shall charge to $B$ a subset of the edges in $E(B, B')$, as described next.

Recall first that  each part in the partition $\calP$ of $G[\light]$
has size at most $\sz^2\wtd^2$.
Let $E^b(B,B')$ denote the subset of edges in $E(B,B')$
  whose rank is in the bottom $(\eps/(8c(h)))\cdot |E(B,B')|$ edges of $E(B,B')$.
We  charge all the edges in $E^b(B,B')$ to $B$. The rationale is that for these edges it is likely that the algorithm won't sample an edge in $E(B,B')$ with lower rank.
The total number of such edges is at most $(\eps/(8c(h)))\cdot |E| \leq (\eps/8)n$.
Let $u$ be the leader of the cluster $B$, and let
$N^b(u,B')$ be the set of vertices, $y \in N(u)$, such that:
$$\left(y\in B' \mbox{ and } (u, y)\in E^b(B,B')\right)
     \mbox{ or } \left(\exists (y',z)\in E^b(B,B') \mbox{ s.t. } y'\in {\cal P}(y)\right)\;.$$
That is, $N^b(u,B')$ is the subset of neighbors of $u$ such that if
Algorithm~\ref{alg:ssg}
selects one of them in Step~\ref{step:sam}, then it obtains an edge in $E^b(B,B')$.
We consider two cases.

\paragraph{First case: $|N^b(u,B')|/|N(u)| < \eps^2/(2^7 c^2(h) \sz^2 \wtd^3)$.}
In this case we charge all edges in $E(B,B')$ to $B$.
For each part ${\cal P}(y)$ such that $y \in N^b(u,B')$ there are at most
$|{\cal P}(y)|\cdot \wtd$ edges $(y',z)\in E^b(B,B')$ for $y'\in {\cal P}(y)$.
Therefore, in this case
$|E^b(B,B')| \leq  \sz^2\wtd^3\cdot |N^b(u,B')| \leq N(u) \cdot \eps^2/(2^7 c^2(h))$
and hence $|E(B,B')| < (8c(h)/\eps)\cdot \eps^2/(2^7 c^2(h)) \cdot N(u)$.
It follows that the total number of charged edges of this type is at most
$ (\eps/( 2^4 c(h)))\cdot 2|E| \leq (\eps/8)n$.

\paragraph{Second case: $|N^b(u,B')|/|N(u)| \geq \eps^2/(2^7 c^2(h) \sz^2 \wtd^3)$.}
For each $u$ and $B'$ that fall under this case we define the set of edges $Y(u, B') = \{(u, v): v \in N^b(u,B')\}$ and denote by $Y$ the union of all such sets (over all such pairs $u$ and $B'$).
Edges in $Y$ are charged to $B$ if and only if they belong to $F$ and are incident to a vertex in $B$.
Fix an edge in $Y$ that is incident to $B$, and note that the selection of neighbors of $u$
is done according to a $t$-wise independent distribution for $t > 4q$, where $q$ is the sample size set in
Step~\ref{step:sam} of the algorithm.
Therefore, the probability that the edge belongs to $F$ is
upper bounded by $(1-\eps^2/(2^7 c^2(h) \sz^2 \wtd^3))^q$, which by the setting of $q$, is at most $p=\eps/(8c(h))$.

We next show, using Chebyshev's inequality, that with high probability, the number of edges in $Y$ that are in $F$ is at most $2p|E|$.
For $y \in Y$, define $J_y$  to be an indicator variable that is $1$ if and only if $y\in F$.
Then for a fixed $y\in Y$, $\E[J_y] \leq p$ and $\{J_y\}$ are pairwise independent (this is due to the fact that the samples of every pair of edges are pairwise independent).
Therefore, by Chebyshev's inequality,
$$
\Pr\left[\sum_{y\in Y} J_y \geq 2p|E|\right] \leq \frac{\Var[\sum_{y\in Y} J_y]}{(p|E|)^2} = \frac{\sum_{y\in Y} \Var(J_y)}{(p|E|)^2} \leq \frac{p(1-p)|E|}{(p|E|)^2} =  \frac{1-p}{p|E|} = \frac{1}{\Omega(n)}\;,
$$
and the proof of Claim~\ref{clm:F} is completed.
\EPF

By combining Claims~\ref{clm:Ep-F} and~\ref{clm:F} we get that  $|E'| \leq (1+\eps)\cdot n$ with probability $1-1/\Omega(n)$, as required.

The random seed that Algorithm~\ref{alg:hle} uses consists of two parts.
The first part is for running Algorithm~\ref{alg:ssg}, whose random seed has length $\tilde{O}((h/\eps)d\log n)$, and hence for $d = \wtd$, its length is  $\tilde{O}((h/\eps)^2\log n)$.
The second part is for selecting random neighbors in Step~\ref{step:sam}.  
Since for each edge in $E$ we pick a random sample of
$\tilde{\Theta}((c^3(h) \sz^2 \wtd^3)/\eps^3)= \tilde{O}((h^{11}/\eps^8)) $
neighbors independently, we obtain by Theorem~\ref{thm:twise} that a random seed of length 
$\tilde{O}(h^{11}/\eps^8)\log n)$
is sufficient.

The running time of Algorithm~\ref{alg:hle} is upper bounded by a constant times the running time of Algorithm~\ref{alg:ssg} times the number of edges sampled in Step~\ref{step:sam}. 
which by Theorem~\ref{thm:main} and the setting of $q$ in the algorithm is $\poly(h/\eps)$.
This completes the proof of Theorem~\ref{thm:main1}.

\subsection*{Acknowledgement}
We would like to thank Oded Goldreich for his helpful suggestions.
We would also like to thank the anonymous reviewers of this paper for their comments.

\bibliographystyle{plain}
\bibliography{refs2}

\newpage
\appendix

\section{Tables of Notations and Parameters}\label{sec:notation}
\begin{tabular}{ |p{2cm}||p{13cm}|  }
 \hline \hline
 \multicolumn{2}{|c|}{ \vspace{.01ex} \large\bf Table of Notations and Parameters for Section~\ref{sec:centers} }  \\
 \hline\hline\vspace{.01ex}
 {\bf Notation}\vspace{.01ex}  & \vspace{.01ex} {\bf Meaning} \vspace{.01ex} \\
 \hline\vspace{.01ex}
$\Gamma_r(v,G)$ \vspace{.01ex} & \vspace{.01ex} the set of vertices at distance at most $r$ from $v$ (distance-$r$ neighborhood of $v$). \vspace{.01ex} \\ \hline \vspace{.01ex}
$C_r(v,G)$ \vspace{.01ex} & \vspace{.01ex} the subgraph of $G$ induced by $\Gamma_r(v,G)$. \vspace{.01ex} \\ \hline \vspace{.01ex}
$n_r(G)$ \vspace{.01ex} & \vspace{.01ex}  $\max_{v\in V} |\Gamma_r(v,G)|$. \vspace{.01ex} \\ \hline \vspace{.01ex}
$h_s(G)$ \vspace{.01ex} & \vspace{.01ex} maximum $h$ such that $G$ is an $(s, h)$-vertex expander. \vspace{.01ex} \\ \hline \vspace{.01ex}
$s = s(n,\eps)$ \vspace{.01ex}& \vspace{.1ex} $2\sqrt{2n/\eps} \ln n$. \vspace{.01ex} \\ \hline \vspace{.1ex}
 $\ell = \ell(n, \eps)$ \vspace{.01ex} & \vspace{.01ex}\vspace{.1ex} $\sqrt{\eps n/2}$  (number of centers). \vspace{.01ex} \\ \hline \vspace{.01ex}
$L^i_j$ \vspace{.01ex} & \vspace{.01ex}vertices in the $i^{\rm th}$ level of the BFS tree of the center $\z_j$.\vspace{.01ex} \\ \hline \vspace{.01ex}
$S(\z_j)$ \vspace{.01ex} & \vspace{.01ex}vertices that are assigned to the center $\z_j$. \vspace{.01ex}\\ \hline \vspace{.1ex}
$E'(\z)$ \vspace{.01ex} & \vspace{.01ex}edges of the BFS-tree that spans the subgraph induced on $S(\z)$. \vspace{.01ex}\\ \hline \vspace{.01ex}
$P(\z,\z')$ \vspace{.01ex} & \vspace{.01ex}the shortest path between $\z$ and $\z'$ that has minimum lexicographic order among all shortest paths.
\vspace{.01ex}\\ \hline \vspace{.01ex}
$e(\z,\z')$\vspace{.01ex} & \vspace{.02ex}if all vertices on
$P(\z,\z')$ belong either to $S(\z)$ or to $S(\z'$, then $e(\z,\z')$ is
the single edge $(x,y) \in   P(\z,\z')$ such that $x \in S(\z)$ and $y \in S(\z')$. \vspace{.01ex}\\ \hline \vspace{.01ex}
$\tr_{\eps, d}(G)$ \vspace{.01ex}& \vspace{.01ex} the minimum distance $\tr$ ensuring that all but an
$\eps/(2d)$-fraction of the vertices have at least $s$ vertices in their $\tr$-neighborhood. \vspace{.01ex}\\ \hline \vspace{.01ex}
$\tr'_{\eps, d}(G)$ \vspace{.01ex}& \vspace{.01ex}the minimum distance $\tr$ ensuring that all but an
$\eps/(4d)$-fraction of the vertices have at least $s$ vertices in their $\tr$-neighborhood. \vspace{.01ex}\\ \hline
\end{tabular}

\newpage
\begin{tabular}{ |p{2cm}||p{13cm}|  }
 \hline \hline
 \multicolumn{2}{|c|}{ \vspace{.005ex} \large\bf Table of Notations and Parameters for Section~\ref{sec:centers} }  \\
 \hline\hline\vspace{.005ex}
 {\bf Notation}\vspace{.005ex}  & \vspace{.005ex} {\bf Meaning} \vspace{.005ex} \\
 \hline
$\calP$  &    the partition of the graph which the spanning graph construction is based on. \\ \hline
 $\gamma$   &    the probability of a vertex to be a center. \\ \hline
 $k$   &   $\Theta( (\log^2(1/\gamma) + \log d)/\gamma)$ -- size parameter used to bound number of vertices in each part of $\calP$  \\ \hline
$\W$  &   the set of centers.  \\ \hline
$t$  &   $2kd$ (the random string used by the algorithm is $t$-wise independent). \\ \hline
 $\cc(v)$  &   subset of vertices in the connected component of $v$ in $G$. \\ \hline
$Z$  &   the set of vertices $v$ such that $|\cc(v)| \geq k$. \\ \hline
$\cent(v)$  &   the {\em center\/} of $v$ with respect to $\W$.\\ \hline
$\cell(w)$  &    $ \{v\in V: \cent(v) = w \}$ (the {\em cell\/} of $w$ with respect to $\W$). \\ \hline
$\ell_k(v)$  &  the minimum integer $\ell$ such that the BFS tree rooted at $v$ of depth $\ell$ has size at least $k$. \\ \hline
$B_k(v)$  &  vertices in the BFS-tree rooted at $v$ of depth $\ell_k(v)$. \\ \hline
$R$  &   the set of vertices $v$ for which $B_k(v)$ does not contain a vertex (center) in $\W$. \\ \hline
$\calT(w)$  &  the BFS tree rooted at $w$ of the subgraph induced by $\cell(w) \setminus R$ \\ \hline
$\calP'$  &  the partition $\calP$ when restricted to $Z$. \\ \hline
$Y_k(v)$  &   $\{u\in V: v \in B_k(u) \}$ \\ \hline
$\chiW$  &   the characteristic vector of the set $\W$. \\ \hline
$\chiW(S)$  &   the projection of $\chiW$ onto $S$ (where $S$ is a subset of $V$). \\ \hline
$\zeta_v$  &   the random variable indicating whether $v$ is a remote vertex with respect to $\W$. \\ \hline
 $S_v $  &   $\{u\in Z: B_k(u) \cap B_k(v) \neq \emptyset\}$. \\ \hline
$\wtd$  &   $8(c(h))^2/\eps$. \\ \hline
$\heavy$  &   the set of {\em heavy} vertices (i.e., with degree greater than $\wtd$). \\ \hline
$\light$  &   $V\setminus \heavy$, the set of {\em light} vertices. \\ \hline
$\calC(u)$  &   the {\em cluster} of $u$, includes $u$ and all the vertices in the parts that are assigned to $u$. \\ \hline
$F$  &   the subset of edges in $E'\cap (\heavy \times \light)$ that are incident to two different clusters (namely, each endpoint belongs to a different cluster). \\ \hline
$E^b(B,B')$  &  the subset of edges in $E(B,B')$
  whose rank is in the bottom $(\eps/(8c(h)))\cdot |E(B,B')|$ edges of $E(B,B')$. \\ \hline
$N^b(u,B')$  &   for $u$ that is the leader of a cluster $B$, it is the set of vertices in $N(u)$ such that:
$\left(y\in B' \mbox{ and } (u, y)\in E^b(B,B')\right)$
or $\left(\exists (y',z)\in E^b(B,B') \mbox{ s.t. } y'\in {\cal P}(y)\right)\;$\\ \hline
\end{tabular}

\section{Proof of Theorem~\ref{thm:lb}}\label{sec:lowerbound}
\BPFOF{Theorem~\ref{thm:lb}}
Let $V$ be a set of vertices and let $v_0$ and $v_1$ be a pair of distinct vertices in $V$.
In order to prove the lower bound we construct two families of random $d$-regular graphs over $V$, $\calF^+_{(v_0, v_1)}$ and $\calF^-_{(v_0, v_1)}$.
$\calF^+_{(v_0, v_1)}$ is the family of $d$-regular graphs, $G=(V, E)$, for which $(v_0, v_1) \in E$.
$\calF^-_{(v_0, v_1)}$ is the family of $d$-regular graphs for which $(v_0, v_1) \in E$ and
the removal of $(v_0, v_1)$ leaves the graph with two
connected components,\footnote{Although a graph that is drawn uniformly from $\calF^+_{(v_0, v_1)}$ (or $\calF^-_{(v_0, v_1)}$) might be disconnected, this event happens with negligible probability~\cite{Bol01}.
Hence, the proof of the lower bound remains valid even if we consider $\calF^+_{(v_0, v_1)}\cap \calC$ and $\calF^-_{(v_0, v_1)}\cap\calC$ where $\calC$ is the family of connected graphs.}
each of size
$n/2$.
We prove that given $(v_0,v_1)$, any algorithm that performs at most $\sqrt{n}/c$ queries for some sufficiently large constant $c > 1$
cannot distinguish the case in which the graph is drawn uniformly at random from $\calF^+_{(v_0, v_1)}$ from the case in which the graph is drawn uniformly at random from $\calF^-_{(v_0, v_1)}$.
Essentially, if the number of queries is at most $\sqrt{n}/c$, then with high constant probability,
each new query to the graph returns a  new random vertex in both families. By ``new vertex'' we mean a vertex that neither  appeared in the query history nor in the answers history.
Since the algorithm must answer consistently with a connected graph $G'$, for every
graph in the support of $\calF^-_{(v_0, v_1)}$ it must answer positively on the query
$(v_0,v_1)$, with probability $1$ (or with high constant probability in case we allow $G'$ to be disconnected with some small constant probability). But since the distributions on query-answer histories in both
cases are very close statistically, this can be shown to imply that there exist
graphs for which the algorithm answers positively on a large fraction of the edges.

We consider a pair of oracles that generate a random graph from $\calF^+_{(v_0, v_1)}$ and $\calF^-_{(v_0, v_1)}$, respectively, on the fly.
On query $(w,i)$ the oracle returns $(v, j)$ where $v$ is $i$-th neighbor of $w$, denoted $N(w,i)$, and $w$ is the $j$-th neighbor of $v$.
If $w$ does not have an $i$-th neighbor then the oracle returns $\emptyset$.
Both oracles construct the graph on demand, i.e. on on query $(w,i)$, if $N(w,i)$ was not determined in a previous step then the oracle determines $N(w,i)$.

\paragraph{First Oracle}
Recall that a random $d$-regular graph can be generated as follows.
Define a matrix $M = [n] \times [d]$ where the $i$-th row corresponds to the $i$-th vertex.
Partition the cells of $M$ into $nd/2$ pairs, namely, find a random perfect matching on the cells of $M$.
Now let $(w,i)$ and $(v, j)$ be a pair of matched cells, then in the corresponding graph it holds that $N(w,i) = v$ and $N(v,j) = w$.
Thus, to obtain a random $d$-regular graph from the family of $d$-regular graphs $G=(V, E)$, for which $(v_0, v_1) \in E$ proceed as follows.
\BE
\item Match a random cell in $v_0$'s row to a random cell with $v_1$'s row.
\item Find a random perfect matching of the remaining cells.
\EE

Our first oracle, $\mathcal{O}^+_{(v_0, v_1)}$, determines the perfect matching on demand as we describe next.
The oracle keeps a matrix, $M$, with $n$ rows which correspond to the $n$ vertices and $d$ columns, which correspond to $d$ neighbors.
At the initialization step:
\BE
\item All the cells are initialized to $\emptyset$.
\item A single edge is determined: $M(v_0,t_0) := (v_1, t_1)$ and $M(v_1,t_1) := (v_0, t_0)$ where $t_0$ and $t_1$ are chosen uniformly and independently over $[d]$.
\EE
On query $(w,i)$, the oracle $\mathcal{O}^+_{(v_0, v_1)}$, proceeds as follows:
\BE
\item Checks if $M(w,i)$ was determined in a previous step, i.e. if $M(w,i) \neq \emptyset$, if so it returns the matched cell $(u, j)$.
\item Otherwise, it picks uniformly one of the empty cell in the matrix, $(u, j)$.
\item Sets  $M(w,i) := (u, j)$, $M(u, j) := (w, i)$. Namely, adds the edge $(u,w)$ to the graph.
\item Returns $(u,j)$.
\EE
Without loss of generality assume that $n\cdot d$ is even~\footnote{In case that $d\cdot n$ is odd we add an extra entry to the matrix, $(0,0)$, so that $M(w, i) = (0, 0)$ means that $w$ does not have an $i$-th neighbor.}.
Thus, if the oracle is queried on all the entries of the matrix, then the resulting graph is a random $d$-regular graph that contain the edge $(v_0, v_1)$ (where self-loops and parallel edges are allowed).
Notice that the resulting graph is drawn uniformly (and independently of the order of the queries) from the family of graphs $\calF^+_{(v_0, v_1)}$.

\paragraph{Second Oracle}
Our second oracle, $\mathcal{O}^-_{(v_0, v_1)}$ generates a random graph from the family of graphs $\calF^-_{(v_0, v_1)}$.
The oracle $\mathcal{O}^-_{(v_0, v_1)}$ keeps a pair of matrices $M_0, M_1$.
The matrix, $M_0$, has $\lfloor n/2 \rfloor$ rows and $d$ columns and the matrix, $M_1$, has $\lceil n/2 \rceil$ rows and $d$ columns~\footnote{Here too, if the number of entries in a matrix is odd we add an entry $(0,0)$.}.
At the initialization step:
\BE
\item All the cells of both matrices are initialized to $\emptyset$.
\item The rows of the matrices are not allocated to any vertex.
\item A random row in $M_b$, $i_b$, is allocated to $v_b$ for each $b\in\{0,1\}$.
\item A single edge is determined: $M_0(i_0,t_0) := M_1(i_1, t_1)$ and $M_1(i_1,t_1) := M_0(i_0, t_0)$ where $t_0$ and $t_1$ are chosen uniformly and independently over $[d]$.
\EE
For a vertex $v$, let $r(v)$ denote the index of the row that is allocated to $v$ and the corresponding matrix by $M_v$.
Given a row $j$ and a matrix $M$, let $v(M, j)$ denote the vertex that the $j$-th row in $M$ is allocated to.
On query $(w,i)$, $\mathcal{O}^-_{(v_0, v_1)}$ proceeds as follows:
\BE
\item If a row was not allocated to $w$ in previous steps then a random row is picked uniformly (from the set of rows that are free) and is allocated to $w$.
\item If the $i$-th neighbor of $w$ was determined previously, i.e., the cell $M_w(r(w), i)$ was matched in previous steps to another cell $M_w(\ell, j)$, then return $(v(M_w, \ell), j)$.
\item Otherwise, select randomly and uniformly an empty cell in $M_w$, $(\ell, j)$. If the $\ell$-th row of $M_w$ is free, then allocate this row to a vertex which is picked uniformly from the set of vertices without an allocated row.
\item Add an edge between $(w, i)$ and $(v(M_w,\ell), j)$, i.e., set $M_w(r(w), i) := M_w(\ell, j)$ and $M_w(\ell, j) = M_w(r(w), i)$. Return $(v(M_w,\ell), j)$.
\EE

Let $\calA$ be an algorithm that interacts with $\mathcal{O}^+_{(v_0, v_1)}$ and let $\pi^+_r = (q^+_1, a^+_1, \ldots, q^+_r, a^+_r)$ be the random variable describing the communication between $\calA$ and  $\mathcal{O}^+_{(v_0, v_1)}$. Namely, $\pi^+_r$ is a list of $r$ queries of $\calA$ and $r$ corresponding answers of $\mathcal{O}^+_{(v_0, v_1)}$. Similarly, define $\pi^-_r$ for $\mathcal{O}^-_{(v_0, v_1)}$.
We claim that for $r = c\sqrt{n}$, the distribution of $\pi^+_r$ is statistically close to the distribution of $\pi^-_r$.
Recall that every query $q_i$ is a pair $(q_{i,1}, q_{i,2})$ where the first entry denotes a name of a vertex and the second entry denotes an index in $[d]$.
Similarly, every answer $a_i$ is a pair $(a_{i,1}, a_{i,2})$.
Let $A^+$ be the event that for every $i\in [r]$ it holds that, $a^+_{i,1} \neq a^+_{j,1}$ for every $i\neq j\in [r]$ and $a^+_{i,1} \neq q^+_{j,1}$ for every $j \in [r]$. Similarly, define $A^-$.
Without loss of generality we assume that $\calA$ does not make queries that does not reveal any new information on the graph.
Hence, in words, $A^+$ (respectively, $A^-$) is the event that whenever $\mathcal{O}^+_{(v_0, v_1)}$ (respectively $\mathcal{O}^+_{(v_0, v_1)}$) selects an empty (un-matched) cell, the corresponding row is empty as well. It is easy to verify that the distribution of $\pi^+_r$ conditioned on $A^+$ is identical to the distribution of $\pi^-_r$ conditioned on $A^-$.
Hence, the statistical distance between the distribution of $\pi^+_r$ and the distribution of $\pi^-_r$ is bounded by $2|\Pr(A^+) -\Pr(A^-)|$.
Since both $\Pr(A^+)$ and $\Pr(A^-)$ are upper bounded by $2r \cdot \frac{2r}{n/2} = 8r^2/ n = 8c^2$, we obtain that the statistical distance is bounded by $16c^2 = 16/49< 1/3$ for $c=1/7$.

From this the lower bound is implied as follows.
Assume towards contradiction that there exists an LSSG algorithm, $\mathcal{A}$, with query complexity $c\sqrt{n}$.
When $\mathcal{A}$ interacts with $\mathcal{O}^-_{(v_0, v_1)}$ it must answer yes with probability $1$ (over the random coins of $\mathcal{A}$ and $\mathcal{O}^-_{(v_0, v_1)}$) when queried on $(v_0, v_1)$.
Since $\mathcal{O}^-_{(v_0, v_1)}$ and $\mathcal{O}^+_{(v_0, v_1)}$ are statistically close, $\mathcal{A}$ that interacts with $\mathcal{O}^+_{(v_0, v_1)}$ must answer yes with probability at least $2/3$ (over the random coins of $\mathcal{O}^-_{(v_0, v_1)}$ and for every setting of random coins of $\mathcal{A}$) when queried on $(v_0, v_1)$.
This implies that for every setting of random coins of $\mathcal{A}$, $\mathcal{A}$ answers yes on at least $2/3$ of the $d$-regular graphs that contain $(v_0, v_1)$.
By an averaging argument it follows that there exists a $d$-regular graph, $G$, such that $\mathcal{A}$ answers yes on at least $2/3$ of the edges of $G$.
A contradiction.

Now consider the setting in which we allow $G'$ to be disconnected with probability $1-\delta_1$ and we require that $|E'| \leq (1/3)nd$ with probability at least $1-\delta_2$ for some constants $0 < \delta_1, \delta_2 \leq 1$.
In this case $\mathcal{A}$ must answer yes with probability $1-\delta_1$, over the random coins of $\mathcal{A}$, when queried on $(v_0, v_1)$ and interacting with $\mathcal{O}^-_{(v_0, v_1)}$ (for any setting of random coins of $\mathcal{O}^-_{(v_0, v_1)}$).
Therefore, by an averaging argument, there exists a $d$-regular graph $G$, such that in expectation, $\mathcal{A}$ answers yes on at least $(2/3)(1-\delta_1)$ of the edges of $G$.
However, since the expected size of $E'$ is bounded by $(1-\delta_2)(1/3) n d + \delta_2 n d$ we reach a contradiction whenever $\delta_1 + \delta_2 < 1/2$.
\EPFOF

\section{A non-polynomial relation between $k$ and $Y_k(v)$}\label{app:Y-k-big}
Recall that $Y_k(v) \eqdef \{u\in V: v \in B_k(u) \}$.
In the following lemma we show that $|Y_k(v)|$ can be super polynomial.
\BL\label{lem:Y-k-big}
There exists a graph $G=(V,E)$ with degree bounded by $d$ and $v\in V$ such that
$|Y_k(v)| = k^{\Omega(\log \log d)}$.
\EL
\BPF
The graph $G$ is a tree, rooted at $v$, and defined as follows.
For simplicity, we let the degree bound be $d+1$ (so that a vertex may have $d$ children, and
hence degree $d+1$).
We partition the levels of the tree into consecutive subsets: $L_0 = \{1,\dots,\ell_0\}$ (where the
root is at level $1$),
$L_1 = \{\ell_0+1,\dots,\ell_1\}$, $\dots$, $L_r = \{\ell_{r-1}+1,\dots,\ell_r\}$.
For each subset $L_i$, and for each level $j$ in the subset, all vertices in level $j$ have the same number of children, which is $d_i \eqdef d^{2^{-i}}$. We set $r = \log\log d$, so that all vertices in
levels belonging to $L_r$ have two children.
Finally we set $s_i = |L_i| = \log_{d_i} g^{1/(r+1)}$, where $g$  determines the size of the
tree, as well as the minimum $k$ that ensures that all vertices in the tree belong to $Y_k(v)$.

By the construction of the tree, the number of vertices in it is of the order of
$\prod_{i=0}^r d_i^{s_i} = g$. In order to upper-bound $k$ (such that all vertices belong to
$Y_k(v)$), consider any vertex $u$ in some level $j \in L_i$, where $0 \leq i \leq r$.
Since $d_i = d_{i-1}^{1/2}$, so that $s_t = 2 s_{t-1}$ for each $t$, we get that $s_i \leq \sum_{i' < i} s_{i'}$. Therefore,
$\d(u,v) \leq 2s_i$. It follows that the number of vertices in the subtree rooted at $u$
that are at distance at most $\d(u,v)$ from $u$ is upper bounded by
$d_i^{s_i}\cdot d_{i+1}^{s_i} < g^{3/2(r+1)}$. Since this is true for every vertex in the tree, we
get that $|\Gamma_{\d(u,v)}(u)| = O(g^{3/2(r+1)}\cdot s_r) = O(g^{3/2(r+1)}\cdot\log g)$,
which gives us an upper bound on $k$, from which the lemma follows.
\EPF

\end{document}